\documentclass[a4paper,11pt]{article}
\pdfoutput=1 % if your are submitting a pdflatex (i.e. if you have
\usepackage{jheppub} % for details on the use of the package, please
\usepackage{graphicx}				% Use pdf, png, jpg, or eps§ with pdflatex; use eps in DVI mode
\usepackage{enumitem}								% TeX will automatically convert eps --> pdf in pdflatex		
\usepackage{amssymb}
\usepackage{amssymb}
\usepackage{caption}
\usepackage{amsmath}
\usepackage{multirow}
\usepackage{makecell}
\usepackage[toc,page]{appendix}
\usepackage{hyperref}
\usepackage{listings}
\usepackage{xcolor}
\usepackage{framed}
\usepackage{array}
\usepackage{float}
\usepackage{amssymb}
\usepackage{breqn}
\usepackage{multicol} % Per gestire le minipage 
\usepackage{booktabs} % Per le linee orizzontali nella tabella
\usepackage{tabularx} 

\usepackage[absolute,overlay]{textpos}
\usepackage[italian, english]{babel}
\usepackage[T1]{fontenc} % if needed
\usepackage{xcolor}
\usepackage{comment}
\usepackage{physics} % Physics package for various mathematical commands
\usepackage{siunitx} % siunitx for consistent unit handling
\AtBeginDocument{\RenewCommandCopy\qty\SI}
\usepackage{graphicx}
\usepackage{subcaption}

\newcommand{\AddrUnina}{Dipartimento di Fisica E. Pancini, 
Universit\`a di Napoli Federico II \\
and INFN, Sezione di Napoli \\
Complesso Universitario di Monte Sant'Angelo, Via Cinthia, Napoli (NA), Italy.}
 \newcommand{\AddrLNGS}{INFN, Laboratori Nazionali del Gran Sasso,
67100 Assergi, L’Aquila (AQ), Italy.}

\newcommand*{\blu}{\textcolor{blue}}

\title{%\red{The time structure of neutrinos from SN1987A}
{The flux of electron antineutrinos from supernova SN1987A data}}

\author[a]{Riccardo Maria Bozza,}
\author[a]{Vigilante di Risi,}
\author[a]{Giuseppe Matteucci,}
\author[a]{Veronica Oliviero,}
\author[a]{Giulia Ricciardi,}
\author[b]{Francesco Vissani}

\affiliation[a]{\AddrUnina}
\affiliation[b]{\AddrLNGS}
\emailAdd{riccardomaria.bozza@unina.it}
\emailAdd{vigilante.dirisi@unina.it}
\emailAdd{giuseppe.matteucci@unina.it}
\emailAdd{veronica.oliviero2@unina.it}
\emailAdd{giulia.ricciardi2@unina.it}
\emailAdd{francesco.vissani@lngs.infn.it}
\abstract{By adopting a state-of-the-art parameterized model of electron antineutrino emission, we have 
   made  some steps  forward
  in the analysis of the thermodynamical properties and  temporal structure of neutrino emission from core collapse SN1987A.
   Our analysis,  unlike similar previous ones,  takes into account the times, energies and angles of arrival of the detected events 
  in a reliable  framework  which includes a  finite ramp in the initial stage of the neutrino emission.
The existence of the accretion phase is confirmed with a confidence level of about 99\%, and the neutrons involved in the emission during accretion, for the first time, do not show significant tensions with expectations, being %about few per cent of the total mass in .
 about 0.02 $M_\odot$.
We determine the parameters of the cooling emission and discuss its duration,  that compares well with theoretical expectations.
We   estimate the delay times between the first neutrino and the first event in the detectors. The goodness of fit checks, performed on the temporal, energy and angular distributions, show that the flux resulting from the best-fit analysis is  compatible with the observed data.}
\keywords{Neutrino emission, Supernova, SN1987A, neutrino flux}
\begin{document}
\maketitle
\parskip0.4ex

\section{Introduction}
%\blu{\begin{itemize}

It is generally accepted that the role played by neutrinos 
in core collapse supernovae (SN)
is crucial, both for the energy transport  
 and the formation of a neutron star \cite{Colgate1966ApJ...143..626C,Wilson1971,Nadyozhin:1978zz, Bethe:1985sox, 1988ApJ...334..891B,1990PhT....43i..24B,Woosley_2005,2007PhR...442...38J,Raffelt:2012kt,Janka:2012wk,RevModPhys.85.245,Foglizzo_2015,Mirizzi:2015eza,Horiuchi_2018,Janka_2017,Janka:2017vlw,7097756f985a410398cc602775095da9, Mezzacappa:2020oyq, Burrows:2020qrp, woosley1986physics}. 
In 1987, three neutrino detectors recorded  a few dozen of events
%some dozens of events 
\cite{hirataPhysRevLett.58.1490,PhysRevD.38.448,PhysRevLett.58.1494,PhysRevD.37.3361,ALEXEYEV1988209}
 which appeared simultaneous, within the limits of their %wide 
 time uncertainties.
These were the first neutrinos that could be attributed with certainty to an extrasolar source, the supernova SN1987A. After so long, they remain the only detected supernova neutrinos. %In anticipation of   the much higher number   expected from a future nearby supernova,
Their significance must be assessed as accurately as possible, in order to offer directions for the analysis of the much higher number  of neutrinos expected from a future (nearby) supernova. 
%Although the complex dynamics of core collapse supernova (SN) is not fully understood,
A particular relevant and interesting aspect, 
indicated by 
the most detailed analyses of the SN1987A events \cite{Loredo:2001rx,Pagliaroli:2008ur}, is the evidence of two distinct phases of neutrino emission.
%as revealed by 

%It is very  that the most refined analyses of these events have obtained an indication of two distinct phases of neutrino emission .
%
%Taking in mind these considerations,

In this paper, we proceed to analyze the SN1987 neutrino events in a time window of $30\,\mbox{s}$ taking into account time, energy and scattering angle of the events (see table~\ref{tab:tab0}), with the aim of extracting the maximum information with
%With these considerations in mind, we analyze 
%the observed events and extract the maximum amount of information through
a state-of-the-art statistical and theoretical approach.
We improve on earlier reference analyses
\cite{Abbott:1987bm, Loredo:2001rx,Pagliaroli:2008ur,DedinNeto:2023hhp} by using a model for SN1987A neutrino emission  that, in consistence with simulations, includes a finite signal rise time \cite{sym13101851}. 
We perform a statistical analysis which is in many ways   more refined and accurate than previous ones, 
%We compare the expectations and observations using
%By comparing expectations and observations 
%with the aid of a likelihood function,  
and estimate the physically motivated parameters of the model together with their uncertainty intervals. 
In this way, we obtain new quantitative and more robust results on the temporal and energy structure of the neutrino emission, and in particular on the duration and intensity of the two main emission phases, and 
on the delay times between the first neutrino and the first event in the detectors.
We  estimate the neutrons involved in the emission during accretion, which for the first time 
do not exhibit significant discrepancies with expectations, 
being between 0.01 and 0.04 $M_\odot$.
Finally, we evaluate the goodness of fit of the model indicated by the analysis.

\begin{comment}
Neutrinos emitted from the supernova  SN1987A were the first neutrinos created outside our solar system to be detected. 
It remains  the only case of a measured neutrino signal from stellar core collapse. 
As of today, there is a recognized need to revisit the data collected from SN1987A, 
as  much higher statistical  data  will be provided in  a future supernova (SN)  event case by current (and planned) large-scale detectors. Another reason are  advancements in theoretical models and simulations, which have improved upon the original analysis
of SN1987A. In particular, today appear consolidated  the presence of both accretion and cooling phases in the SN evolution.

Under these premises, we analyze the neutrino events of SN1987A to extract the maximum amount of information through a comprehensive statistical approach. By reconstructing the likelihood function for all detected events, we optimize this probability by varying key theoretical parameters associated with the emission by using a parameterized model \cite{sym13101851}. Our analysis focus on  the time-energy distribution
of the signal, an aspect often neglected in previous analyses.
\end{comment}

%\section{Context and experiments}
%\section{Neutrino detectors}
%\paragraph{Kamiokande-II}
%\paragraph{IMB}
%\paragraph{Baksan}

%\section{{Expectations}}
%\subsection
\section{Neutrino emission in core collapse supernovae}
\label{s21}

Up to date, the only   neutrino data from a core collapse supernova (SN) available  are the neutrino events following  SN1987A,  
that exploded in the Magellanic Cloud at a distance of \( D = 51.4 \pm 1.2 \) kpc \cite{Panagia}. \footnote{See e.g. \cite{woosley1986physics,Roulet:2022akk, Ricciardi:2024dka}   for historical and didactic descriptions of SN physics.} They were collected by Kamiokande-II, the Irvine-Michigan-Brookhaven detector (IMB) and the Baksan Neutrino Observatory (BNO) in three mutually compatible time windows \cite{hirataPhysRevLett.58.1490,PhysRevD.38.448,PhysRevLett.58.1494,PhysRevD.37.3361,ALEXEYEV1988209}. \footnote{It  would be interesting to include the two events that LSD has observed in the 30\,s time window which includes the neutrino burst recorded by the other detectors; however, from the existing literature   \cite{Badino:1984ww,Aglietta:1987it,Aglietta:1987we,Dadykin:1987ek,Saavedra:1987tz}, we were unable to find the information on the background, that we need to perform the analysis.}
%
%supports  the existence of an emission phase,  where a large non-thermal neutrino emission starts
%\cite{Loredo:2001rx, Pagliaroli:2008ur}. %\footnote{Details on the neutrino production in the accreting region and their role in reviving the explosion can be found in \cite{Janka:2000bt} and the Appendix of \cite{Pagliaroli:2008ur}.}.
%
Although SNs produce neutrinos of all flavours, these detectors  were practically  sensitive only to electronic antineutrinos, which are the focus of our analysis.

\subsection{Accretion and cooling}
From the beginning all the relevant simulations \cite{ Colgate1966ApJ...143..626C,Wilson1971,Nadyozhin:1978zz, Bethe:1985sox, 1988ApJ...334..891B,1990PhT....43i..24B,Woosley_2005,2007PhR...442...38J,Raffelt:2012kt,Janka:2012wk,RevModPhys.85.245,Foglizzo_2015,Mirizzi:2015eza,Horiuchi_2018,Janka_2017,Janka:2017vlw,7097756f985a410398cc602775095da9, Mezzacappa:2020oyq, Burrows:2020qrp, woosley1986physics}, based on different theoretical modeling of the SN explosion, have shown the existence of an initial and very intense neutrino emission. This harmonizes with the results of the time-energy analyses of SN1987A data \cite{Loredo:2001rx,Pagliaroli:2009qy,Pagliaroli:2007sda,DedinNeto:2023hhp}: there is  an initial non-thermal neutrino emission during the formation of the neutron star  lasting a fraction of second (accretion), followed by a  longer phase of thermal emission (cooling).
We adopt a  parametric framework \cite{sym13101851} with three stages, which aims to model not only the accretion  and cooling phases, but also  the brief initial phase in which the brightness increases, that will be observable
in future \cite{sym13101851, PhysRevD.80.087301, Pagliaroli:2009qy}. 

The choice of which model to adopt for neutrino emission is crucial for the analyses. In principle, models derived from \emph{ab initio} calculations would be preferable. The problem, however, is that in general %numerical simulations 
the  numerical simulations still  fail to provide an accurate estimate of uncertainty.

\begin{table}[t!]
\centerline{
\begin{tabular}{||c|rrcc||c|rrcc||}
\hline
& \footnotesize Relative\, & \footnotesize Energy &\footnotesize  SN-angle &\footnotesize  Backgr. & &\footnotesize Relative\, & \footnotesize Energy &\footnotesize  SN-angle &\footnotesize  Backgr. \\[-1ex]
&\footnotesize time \tiny [\unit{\ms}]  & \tiny [\unit{\MeV}]\ \ \ &\tiny  [deg] &\tiny  [\unit{\Hz/\MeV}] & &\footnotesize time \tiny [\unit{\ms}]  & \tiny [\unit{\MeV}]\ \ \ &\tiny  [deg] &\tiny  [\unit{\Hz/\MeV}]  \\ \hline
K1 & 0 & 20.0$\pm$2.9 &18$\pm$18 & 1.0E-5 & I1 & 0 & 38$\pm$7 &80$\pm$10 & $10^{-5}$(?)\\
K2 & 107 & 13.5$\pm$3.2 &40$\pm$27 & 5.4E-4 &I2 & 412 & 37$\pm$7 &44$\pm$15 & $10^{-5}$(?)\\
K3 & 303 & 7.5$\pm$2.0 &108$\pm$32 & 2.4E-2 &I3 & 650 & 28$\pm$6 &56$\pm$20 & $10^{-5}$(?)\\
K4 & 324 & 9.2$\pm$2.7 &70$\pm$30 &2.8E-3&I4 & 1141 & 39$\pm$7 &65$\pm$20 &$10^{-5}$(?)\\
K5 & 507 & 12.8$\pm$2.9 &135$\pm$23 & 5.3E-4&I5 & 1562 & 36$\pm$9 & 33$\pm$15& $10^{-5}$(?)\\
K6 & 686& 6.3$\pm$1.7 & 68$\pm$77& 7.9E-2&I6 & 2684& 36$\pm$6 &52$\pm$10 & $10^{-5}$(?)\\
K7 & 1541& 35.4$\pm$8.0 &32$\pm$16 & 5.0E-6&I7 & 5010& 19$\pm$5 &42$\pm$20 & $10^{-5}$(?)\\
K8 & 1728& 21.0$\pm$4.2 &30$\pm$18 & 1.0E-5&I8 & 5582& 22$\pm$5 & 104$\pm$20& $10^{-5}$(?)\\ 
K9 & 1915& 19.8$\pm$3.2&38$\pm$22 & 1.0E-5&&&&&\\
K10 & 9219& 8.6$\pm$2.7&122$\pm$30 & 4.2E-3 &&&&&\\
K11 & 10433& 13.0$\pm$2.6&49$\pm$26 & 4.0E-4&&&&&\\
K12 & 12439& 8.9$\pm$1.9 &91$\pm$39 & 3.2E-3&B1 & 0 & 12.0$\pm$2.4 & 90(?) & 8.4E-4\\
K13 & 17641 & 6.5$\pm$1.6& $103\pm50$ & 7.3E-2&B2 & 435 & 17.9$\pm$3.6 & 90(?) & 1.3E-3\\
K14 & 20257&  5.4$\pm$1.4&  $110\pm50$ &5.3E-2&B3 & 1710 & 23.5$\pm$4.7 & 90(?) & 1.2E-3\\
K15 & 21355& 4.6$\pm$1.3&  $120\pm50$ & 1.8E-2&B4 & 7687 & 17.5$\pm$3.5 & 90(?) & 1.3E-3\\ 
K16 & 23814& 6.5$\pm$1.6&  $112\pm50$ &7.3E-2  & B5 & 9099 & 20.3$\pm$4.1 & 90(?) & 1.3E-3\\ \hline
\end{tabular}}
\caption{\emph{Properties of the events in the neutrino bursts observed  in the occasion of SN1987A. The events are indicated by K1, K2... K16 for Kamiokande-II \cite{hirataPhysRevLett.58.1490,PhysRevD.38.448,Krivoruchenko:1988zg}; I1, I2... I8 for IMB \cite{PhysRevLett.58.1494,PhysRevD.37.3361}; B1, B2 ... B5 for Baksan \cite{ALEXEYEV1988209}. The background in last column is taken from table~1 in \cite{Vissani:2014doa}.
A question mark reminds us that the background rate at the given energies and angles in IMB, and the directions of arrival of events in Baksan are not measured.
We adopt the values in the table, since we have verified that the analysis does not depend critically on them.}
%\red{Aggiungere Refs e spiegare il motivo dei 5 eventi in piu' rispetto al lavoro originale. Inoltre rispetto a (vedi latex)
% https://inspirehep.net/files/dd17d6a098701bf5034274e62186d459
%ci sono differenze arrore energia K12 1.9 e SN angolo K2 15 e K5 137; Forse sto guardando il preprint e i dati sono stati corretti dopo? Urgono Refs.}
}
\label{tab:tab0}
\end{table}
%similar in quality to that which allowed John Bahcall to guide solar neutrino science since the 1960s. 
This state of affairs calls forth the adoption of models which are motivated by  the physics of the problem  and include parameters capable of describing theoretical uncertainties. In this way, data analysis can proceed without relying on hypotheses that are not fully consolidated or based on quantitative details at risk of subsequent reconsideration.
Parametric models with well motivated parameters are also 
%motivated on physical bases and therefore are 
easy to interpret and adaptable to future
 observations.
 
We  improve on  the most common types of %parametric
models
used in data analysis—those that assume quasi-thermal distributions, neglect
the accretion phase altogether or describe it partially.
The presence of a  third stage is required in order to give a smooth behavior of the observable neutrino distributions in the whole time range~\cite{sym13101851}.
This amends models where the accretion phase starts at the luminosity peak, such as in refs.~\cite{Loredo:2001rx,Pagliaroli:2008ur,DedinNeto:2023hhp}.

The $\bar \nu_e$ spectrum,
differential in time and energy, is given,  in each instant of the emission, by the sum of a component of accretion (a) and one of cooling (c):
\begin{equation}
    \dv{\dot{N}_{\nu}}{E_{\nu}}(E_{\nu},t)=\dv{\dot{N}_{\nu,a}}{E_{\nu}}(E_{\nu},t)+\dv{\dot{N}_{\nu,c}}{E_{\nu}}(E_{\nu},t).
\end{equation}
The emitted spectrum of the thermal cooling is given by \cite{sym13101851}
\begin{equation}\label{eq:spectrum_cooling}
\dv{\dot{N}_{\nu,c} }{E_{\nu}}(E_{\nu}, t) = \frac{c}{(hc)^3} \times \pi R^2_{ns} \times \frac{4 \pi E_{\nu}^2}{1 + \exp(E_{\nu}/T_c)} , 
\end{equation}
where $R_{ns}$ is the radius of the nascent proto-neutron star and $T_c$ is its temperature. %\textcolor{blue}{Aggiungere commenti sui valori attesi per Rns e Tc}

%Specifically for electronic antineutrinos, the most relevant reaction occurring in the accreting region is $n + e^+ \rightarrow p + \bar{\nu}_e$.

Let us move to the discussion of emission in the accretion phase. We indicate with  \(T_a\) the temperature of the positrons,
and  $N_n$ the number of neutrons
of the SN outer core %mass 
that participate in antineutrino production via  $e^++n\to p+\bar{\nu}_e$. It is convenient to  express  the latter as $N_n=\xi_n \times M_\odot/m_n$, where $\xi_n$ quantifies the {\em fraction of neutron mass} contained in one solar mass of neutrons ($M_\odot$ is the mass of the Sun, \(m_n\)  the mass of one neutron).  The rate of emission of electron antineutrinos in the accreting region is given by \cite{sym13101851}:
\begin{dmath}
\label{eq:spectrum_accretion}
\dv{\dot{N}_{\nu,a}}{E_{\nu}}(E_{\nu}, t) = \frac{c}{(hc)^3} \times \xi_n \times \frac{M_{\odot}}{m_n}\times\sigma_{e^+ n}(E_{\nu}) \times \frac{ 8 \pi E_e^2}{1 + \exp(E_e/T_a)} \,,
\end{dmath}
where $E_e$ is the positron energy and $\sigma_{e^+ n}(E_v)$ is the cross-section for the process of positron capture. The non-thermality of the spectrum is evident from the dependence on  $\sigma_{e^+ n}(E_v)$. 
The dependence of $E_e$ and  $ \sigma_{e^+ n}$ on the antineutrino energy $E_v$ is given in \cite{Pagliaroli:2008ur}. %Formulas for $E_e$ as function of the antineutrino energy $E_{\nu}$ and  \(\sigma_{e^+ n}(E_v)\) are as in \cite{Pagliaroli:2008ur}. 

%A reasonable assumption is that the luminosities of the accretion and cooling components of the spectrum each have two distinct phases: 
%a first, common one in which the emission increases rapidly, 
%followed by a second phase in which the emission decreases, but this time, each with a different time constant for  accretion and cooling.

\subsection{Time evolution of the $\bar \nu_e$ emission}
\begin{figure}[t!]
    \centering\includegraphics[width=0.9\linewidth]{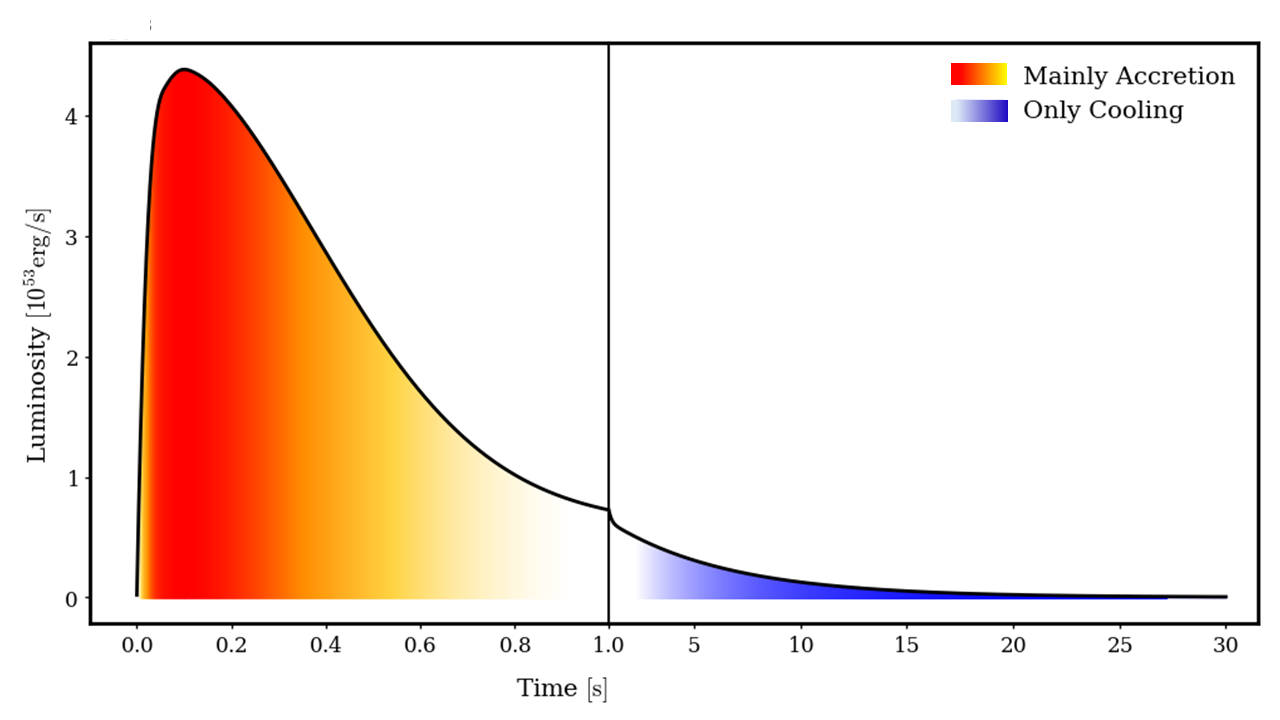}\caption{\emph{Antineutrino luminosity obtained by evaluating our model at the best fit points. The color scale is used to discriminate between the phase which is mainly accretion  (red-orange) and the subsequent only cooling phase (blue). Note the different time units in the left and right panels.
    %The inset in the plot shows the trend during the accretion phase.
    }
     }
    \label{fig:luminosity}
\end{figure}

It is convenient to assume that the luminosities of the accretion and cooling components of the spectrum each have two distinct phases: a   first, common one, characterized by a rapid increase in emission,  followed by a second phase, during which the emission decreases, but  with different time constants $\tau$ for  accretion and cooling.
The two phases are linked in a smooth manner by adopting a function of time $\mathfrak{F}(t)$ which initially grows to $t=t_{\text{max}}$ -- where 
$\mathfrak{F}(t_{\text{max}})=1$ -- 
and then slowly decreases, %, see Eq.~(23) of Ref.
 namely~\cite{sym13101851}:
\begin{equation}\label{eq:Fcalligrafica}
  \mathfrak{F}  (t, t_{\mbox{\tiny max}}, \tau, \alpha)=\bigg(\frac{1+\alpha\big(\frac{t_{\mbox{\tiny max}}}{t}\big)^{\alpha}}{\exp[2(\big(\frac{t}{\tau}\big)^\alpha-\big(\frac{t_{\mbox{\tiny max}}}{\tau}\big)^{\alpha})]+\alpha \big(\frac{t_{\mbox{\tiny max}}}{\tau}\big)^{\alpha}\big(\frac{t_{\mbox{\tiny max}}}{t}\big)^2}\bigg)^{\frac{1}{2}}.
\end{equation}
where the parameter $\alpha$ governs the long-time exponential decrease of the function $\mathfrak{F}$:  $\alpha=2$ for accretion and $\alpha=1$ for cooling.
In this way, we reproduce the temporal characteristics of the emission suggested by simulations in a simple way and we can vary the parameters to account for residual theoretical uncertainties.
The  luminosity resulting from our analysis,  displaying the various phases of electron antineutrino emission, is presented in figure~\ref{fig:luminosity}.
For data analysis  we consider the model advocated in eqs.~(28) and (29) of~\cite{sym13101851},  that we recall here. As the accretion phase proceeds,  we can reasonably assume that the positron temperature $T_a$ does not vary much, while the neutron density $\xi_{n}$ varies in time.
Furthermore, it seems safe to presume that  the radius $R_{ns}$ of the nascent neutron star  stays approximately constant, while its temperature $T_c$ varies in time as it cools down until the end of the emission. \footnote{Let us point out that alternative descriptions of the time evolution of the parameters are also possible. For example, 
  a scenario where the neutron star radius decreases and the cooling temperature remains constant is suggested in \cite{sym13101851}, although deemed less reasonable on  physical grounds.}
%}

The expression for the antineutrino flux is straightforwardly obtained by the emission spectra:
  \begin{equation}
      \Phi_{\bar{\nu}_{e}}(E_{\nu_{e}},t)=\frac{1}{4\pi D^2}\bigg(\dv[]{\dot{N}_{\nu,a}}{E_{\nu}}+\dv[]{\dot{N}_{\nu,c}}{E_{\nu}}\bigg),
  \end{equation}
  where $D$ is the distance of the Earth from the supernova. As anticipated, its variation in time is transparently encoded in the time dependence of the parameters $T_c$ and $\xi_n$, thus allowing for an efficient comparison with the experimental data. We have:
  \begin{eqnarray}
     T_c(t) &=& T_0 \times \sqrt[4]{\mathfrak{F}(t, t_{\text{max}}, \tau_c, 1)}, \nonumber \\
  \xi_n(t) &=& \xi_{n0} \times \mathfrak{F}(t, t_{\text{max}}, \tau_a, 2),   
  \label{cool:dep}
  \end{eqnarray}
 where $T_0$ is the temperature related to the average energy of antineutrinos, the 
number of neutrons, $\xi_{n0}$, parameterizes the intensities of the accretion emission and 
the fourth root accounts for the fact that the luminosity (proportional to $T_c^4$) scales with the function $\mathfrak{F}$.
$T_0$ and $\xi_{n0}$ are also the values at the maximum of the initial ramp, that is when $t=t_{\text{max}}$. The time parameters $\tau_a$ and $\tau_c$ characterize the accretion and the cooling phases, respectively.
 We also set: 
\begin{eqnarray}
R_{ns}(t) &=& R_{ns0}, 
\label{R1}\\
T_a(t) &=& 0.6 \times T_0.
\label{R}
  \end{eqnarray}
  %as expected on the basis of core collapse SN modelings. In particular, the 
  The second condition is determined by smoothly matching the average energies of the antineutrinos in the two phases of the emission.
  %at the beginning, that is $t=0\,\unit{s}$ 

A much debated point is the  role of neutrino oscillations in supernovae. Some aspects have been investigated, like %Smirnov and co-workers have pointed out 
the role of the matter effect 
\cite{Mikheyev:1989dy,Dighe:1999bi}
%\cite{mik89,dig99} %Pantaleone has pointed out that the process has 
and the  non-linear nature of the process \cite{Pantaleone:1992eq}, 
but the overall discussion 
of this plausible possibility 
does not  seem to have reached a conclusion  yet \cite{Volpe:2023met}. 
The application of these theoretical 
representations 
%considerations 
to neutrino emission from SN1987A has also occasioned an intense discussion. 
Several older analyses have suggested that oscillations play an important role \cite{Smirnov:1993ku,Lunardini:2000sw,Minakata:2000rx}, while more recent studies, beginning with \cite{Kachelriess:2001sg, Barger:2002px}, have pointed out that astrophysical uncertainties hinder the ability to draw firm conclusion.
An interesting insight emerged by quantitative analyses of the SN1987A data~\cite{Pagliaroli:2007sda,Pagliaroli:2008ur},  recently reaffirmed~\cite{DedinNeto:2023hhp}, namely that the oscillations do not play an important role in the case of a `normal' mass spectrum, while in the case of a `inverse’ mass spectrum there is a potentially interesting effect during the accretion phase. However, the latter conclusion is subject to a number of uncertainties,  due to the incomplete modeling of the oscillations and  the neutrino emission itself. 
%Fortunately, 
In any case, theoretical expectations, as well as experimental results, seem to point towards the `normal' mass spectrum case~\cite{deSalas:2020pgw,Capozzi:2021fjo,Gonzalez-Garcia:2021dve}. 

In light of these considerations, 
%in our study 
we  feel justified in assuming normal mass ordering and 
 we do not add effects due to neutrino oscillations. To be even more conservative and precise, we can say that the neutrino flux  we obtain from our analysis of SN1987A data is  to be considered as an effective flux,  already including  the effects of neutrino oscillations. Its precise meaning in relation to astrophysics requires progress that is currently unavailable, but which we expect would not  change the conclusions by more than a dozen percentage points.

%\section{{Modelling the detectors and building the likelihood}}
%\subsection
\section{Signal and detector response}

%\section{The construction of the likelihood}

%In this section we outline the procedure to compare the observed events of SN1987A with the antineutrino flux of the model proposed above. 
%\blu{inserire le referenze delle analisi di riferimento ed eventualmente omettere formule che si possono trovare su altri riferimenti}

%\subsection{Interactions}

%It is generally accepted that the SN1987A observations are due to electron antineutrinos \ref{}. Even though the signal could be due to other reactions, 
%The inverse beta decay (IBD)
%\begin{equation}
%    \bar{\nu}_e + p\rightarrow n+e^+
%\end{equation}
%is by far the most relevant channel for supernova neutrino and the hypotheses that the SN1987A signal is entirely due to IBD is very reliable. See Refs.\cite{PhysRevD.36.2283,Vissani:2014doa,Tomas:2003xn,Costantini:2004ry,Vissani:2008wbj} for further discussions. 

\subsection{Inverse beta decay}
The inverse beta decay (IBD)
\begin{equation}
    \bar{\nu}_e + p\rightarrow n+e^+
\end{equation}
is by far the most relevant channel for non-thermal processes and the hypotheses that the SN1987A signal in the accretion region is entirely due to IBD is quite reliable \cite{PhysRevD.36.2283,Vissani:2014doa,Tomas:2003xn,Costantini:2004ry,Vissani:2008wbj}. \footnote{The possibility of other types of events besides IBD is discussed in several works, starting with
\cite{PhysRevD.36.2283}; however, this is  considered less plausible by recent quantitative discussions, such as \cite{Vissani:2014doa}.}
We use all the available data on energy, time and the angle $\theta$  between the antineutrino and the positron provided  by Kamiokande-II \cite{hirataPhysRevLett.58.1490,PhysRevD.38.448},  IMB \cite{PhysRevLett.58.1494,PhysRevD.37.3361} and  Baksan \cite{ALEXEYEV1988209} and 
reported in table~\ref{tab:tab0}. 
%
%\footnote{Complementary analysis where other scenarios are investigated can be found in the literature \cite{}.}
%A time window of $30s$ includes all the events and is used for the  analysis.
%

\begin{comment}
%\subsection{The signal rate}

In the IBD reaction the energy of the incoming antineutrino $E_\nu$ is given in terms of the emitted positron energy $E_e$, and the angle $\theta$ between the antineutrino and the positron, which we indicate as scattering angle for simplicity, by the formula:
\begin{equation}
    E_{\nu}=\frac{E_e+\delta}{1-(E_e-p_e \cos{\theta})/m_p},
\end{equation}
where $\delta=(m_n^2-m_p^2-m_e^2)/(2m_p)$ and $p_e$ is the positron momentum. 

\end{comment}

For an ideal detector, the IBD positron spectrum, triply differential in time $t$, positron energy $E_e$, and $\cos \theta$, is given by
\begin{equation}
    S_e(E_e, \cos{\theta}, t)=N_p \times \Phi_{\bar{\nu}_e}(E_{\nu},t)\times \dv[]{\sigma^{\text{IBD}}}{E_e}\,(E_{\nu},E_e)\times J(E_\nu,\cos{\theta}),
\end{equation}
where $N_p$ is the number of protons of the detector
 and $J(E_{\nu},\,\cos{\theta})$ is the Jacobian of the differential flux.
%, is listed in table~\ref{tab:efficiency} 
%for each experiment. 
% 
For the IBD cross section $\dd \sigma^{\text{IBD}}$ we adopt the most recent and accurate calculation to date~\cite{Ricciardi:2022pru}.

%\subsection{Number of protons}
%\label{sect:Np}

%The number of free protons $N_p$   for each detector is given in   section \ref{sect:Np}.

The number of free protons $N_p$ (hydrogen atoms) in each detector is calculated by means of the formula
\begin{equation}
N_p = n(1 - Y_D) \frac{M_{\text{det}}}{m_{\text{mol}}}
\end{equation}
where \( n \) is the stoichiometric number of hydrogen in the molecules which constitute the detectors,
\( Y_D = 0.0145\% \) is the deuterium abundance (as on Earth, in oceans it is slightly more), \( M_{\text{det}} \) is the mass of the detectors ($1 \mbox{kton} =  10^9 \, \text{g}$) and 
\( m_{\text{mol}} \) is the mass of the molecules that compose the detectors.
Values and results are reported in table \ref{tab:protons}.

% non mi concludere il documento :P - scusa <3

\begin{comment}
We include the jacobian
\begin{equation}
    J(E_{\nu},\,\cos{\theta})= \frac{E_{\nu}^2}{m_p}\frac{p_e}{E_e+\delta}
\end{equation}
to obtain the correct expression of the flux differential in the cosine of the scattering angle.
\end{comment}

\subsection{Modeling the detector response}
\label{sect:Modeling}

The %passage
relation between $S_e$ and the  signal 
$S$, the spectra actually observed by the  detectors, 
 can be described as  a convolution of $S_e$ with the intrinsic efficiency function $\eta(E_e)$ and an ad-hoc Gaussian smearing function $G(E_e-E_i, \sigma(E_e))$, where $E_i$ is the measured value of the positron energy and $\sigma(E_e)$ the %variance:
 standard deviation:
\begin{equation}
 \label{eq:differential_observed_signal}
    S(E_i, \cos{\theta}, t)=\int_{m_e}^{\infty}\zeta(\cos{\theta})\,\eta(E_e)\,G(E_e-E_i,\sigma(E_e))S_e(E_e,\cos{\theta},t)\,\dd E_e.
 \end{equation}

 \begin{table}[tb]%[h!]
\centering
\begin{tabular}{|c|c|c|c|c|c|}
\hline
\textbf{Detector} & \( M_{\text{det}} \, [10^9 \, \text{g}] \) & \textbf{Molecule} & \( n \) & \( m_{\text{mol}} \, [10^{-23} \, \text{g}] \) & \( N_p \, [10^{32}] \) \\ \hline
Kam-II            & 2.14                                    & H\textsubscript{2}O & 2       & 2.9915                              & 1.430                   \\ \hline

IMB              & 6.8                                     & H\textsubscript{2}O & 2       & 2.9915                              & 4.546                   \\ \hline

Baksan           & 0.2                                     & C\textsubscript{9}H\textsubscript{20} & 20 & 21.297                              & 0.188                   \\ \hline
\end{tabular}
\caption{\emph{Mass, molecular composition, stoichiometric number of hydrogen, molecular mass and number of free protons for each detector. The information on masses and composition of the detectors are taken from literature \cite{hirataPhysRevLett.58.1490, CASPER1988463, ALEXEYEV1988209}}}
\label{tab:protons}
\end{table}
Here  $m_e$ is the positron mass and the function $\zeta(\cos{\theta})$ is the angular bias function of the detector, which is provided by the experimental collaboration \cite{hirataPhysRevLett.58.1490,PhysRevD.38.448,PhysRevLett.58.1494,PhysRevD.37.3361,ALEXEYEV1988209}. 
 The total efficiency of the detector is also supplied, but not the efficiency function $\eta(E_e)$. 
 We recover the unknown $\eta(E_e)$ and  $\sigma(E_e)$ by extrapolation,
as described in
section~\ref{Sect:effres}.
%~\cite{Vissani:2014doa}, 
%
%We have listed for each event the explicit value of the energy uncertainty $\sigma(E_e)$ in Tab.\,\ref{tab:tab0}.
%They are consistent with the results in \cite{Vissani:2014doa} 
We have tested minor variations in  the function parametrizations and verified that they do not critically affect the global fits or alter the key conclusions regarding the SN1987A neutrino signal.
%We have verified that the final results of the analysis are stable under minor changes of these analytical expressions.  
%It is worth noting that the analysis results that will be shown below are stable and thus independent of the analytical expressions used for $\eta(E_e)$ and  $\sigma(E_e)$. This was verified using different parameterizations.

The advantage of eq.~\eqref{eq:differential_observed_signal} is that, by setting an appropriate minimal value 
$E_{\min}$ on the observed positron   energy, one can take into account the entire SN1987A dataset and background {\em a posteriori} rather than {\em a priori}. 
 Moreover, the number of total events $N_{\text{tot}}$ can be estimated by integrating the observed signal and background over a wide time window (0 to 30 s), avoiding biases,
 over the entire range for $\cos \theta$ (-1 to 1) and  above the threshold  $E_{\min}$ for the positron energy:
\begin{equation}
    N_{\text{tot}}(E_{\min})=N_{\text{bkg}}+\int_{E_{\min}}^{\infty}\zeta(\cos{\theta})\,\varepsilon(E_e, E_{\min})S_e(E_e, \cos{\theta}, t)\,\dd E_e\,\dd\!\cos{\theta}\,\dd t .
\end{equation}
The information on the total number of events due to background $N_{\text{bkg}}$ is taken from table 2 in ref.~\cite{Vissani:2014doa}. The  function $\varepsilon(E_e, E_{\min})$ is  the total efficiency  of the detectors, which results by integrating  the Gaussian smearing function:
\begin{equation}\label{eq:epsilonexpression}
       \varepsilon(E_e, 
        E_{\min})=\eta(E_e)\times \frac{1+ \text{erf}\left(\frac{E_e-E_{\min}}{\sqrt{2}\sigma(E_e)}\right)}{2},
\end{equation}
where erf is the Gauss error function. The values for $E_{\min}$ used in the analysis are listed in table~\ref{tab:efficiency} and discussed in section \ref{Sect:effres}. 
\begin{comment} We set $E_{\min}=4.5\,\unit{MeV}$  for the total efficiency of Kamiokande-II, which allows to include all SN1987A events in the analysis, rather than excluding some of them \emph{a priori} as background.\footnote{Further discussion on the choice of the thresholds can be found in \cite{Vissani:2014doa}.}
\end{comment}
%
%The function $\sigma(E_e)$ is extrapolated by the energy uncertainties furnished by the experimental collaboration.
%The results are in excellent agreement with data.
We recover the total efficiency and number of events provided by the experimental collaborations \cite{hirataPhysRevLett.58.1490,PhysRevD.38.448,PhysRevLett.58.1494,PhysRevD.37.3361,ALEXEYEV1988209}.
Our procedure is in agreement  with that outlined in~\cite{Jegerlehner:1996kx}, and it is also  adopted in more recent analyses, such as \cite{Vissani:2014doa} and~\cite{DedinNeto:2023hhp}.
Instead, some of the  most careful  analyses of the SN1987A data so far, namely  \cite{Loredo:2001rx,Pagliaroli:2008ur}, adopt a suboptimal procedure for the inclusion of efficiency, which introduces a bias in the interpretation of their results, 
as discussed in \cite{Vissani:2014doa}.

\subsection{The response functions in detail}
\label{Sect:effres}
%{\color{magenta} \Large \bf VA DETTO CHE ABBIAMO VERIFICATO CHE USARE  I VALORI DI JPG NON CAMBIA I RISULTATI}\\
%{\color{blue} INGLOBARE LA TABELLA 
In this section, we provide more details on how we model the response of the three detectors  considered.  %Our approach follows the procedure introduced in, e.g., refs.~\cite{Jegerlehner:1996kx,Vissani:2014doa}, with some refinements discussed in the main text. {\color{magenta} In our formalism, the observed signal is a linear functional of the signal in an ideal detector, as described by the eqs.~\eqref{eq:differential_observed_signal}.}
Our approach refines a procedure already adopted e.g. in refs.~\cite{Jegerlehner:1996kx,Vissani:2014doa}.
Eq.~\eqref{eq:differential_observed_signal} describes a linear functional which correlates  the observed and true energy. 
%As anticipated in section~\ref{sect:Modeling},
For each detector, three  ingredients are needed to build this functional:
%As described by the eqs.~\eqref{eq:differential_observed_signal}, the observed signal is a linear functional of the signal in an ideal detector, and For each detector, two ingredients are especially relevant:
\begin{itemize}
    \item[--] the energy resolution \(\sigma(E_e)\), which  describes the spread between the true positron energy \(E_e\) and the reconstructed energy \(E_i\);
    \item[--] the intrinsic efficiency \(\eta(E_e)\), which  characterizes the efficiency of the detector and the reconstruction process, factoring out resolution effects;
    \item[--] the angular bias \(\zeta(\cos\theta)\), which is needed to take directional inefficiencies into account.
\end{itemize}
%Let us proceed with their detailed description.
\paragraph*{Energy resolution \(\sigma(E_e)\):}Each experiment measures the energy of the emitted positron with finite resolution, modeled by a Gaussian kernel:
\begin{equation}
    G\bigl(E_e-E_i, \sigma(E_e)\bigr)=\frac{1}{\sqrt{2\pi}\,\sigma(E_e)} \,\exp\!\Bigl[-\tfrac{(E_{e} - E_i)^2}{2\,\sigma(E_e)^2}\Bigr].
    \label{eq:kernel}
\end{equation}
We parametrize the energy resolution as
\begin{equation}
    \sigma(E_e) \;=\;\sigma_{\mathrm{stat}} \,\sqrt{\tfrac{E_e}{10\,\mathrm{MeV}}}\;+\;\sigma_{\mathrm{syst}}\,\tfrac{E_e}{10\,\mathrm{MeV}},
    \label{eq:sigma_parameterization}
\end{equation}
where \(\sigma_{\mathrm{stat}}\) and \(\sigma_{\mathrm{syst}}\) are constants fitted to the energy uncertainties of each detector reported in table~\ref{tab:tab0}. 

\begin{figure}
    \centering
    \includegraphics[width=1\linewidth]{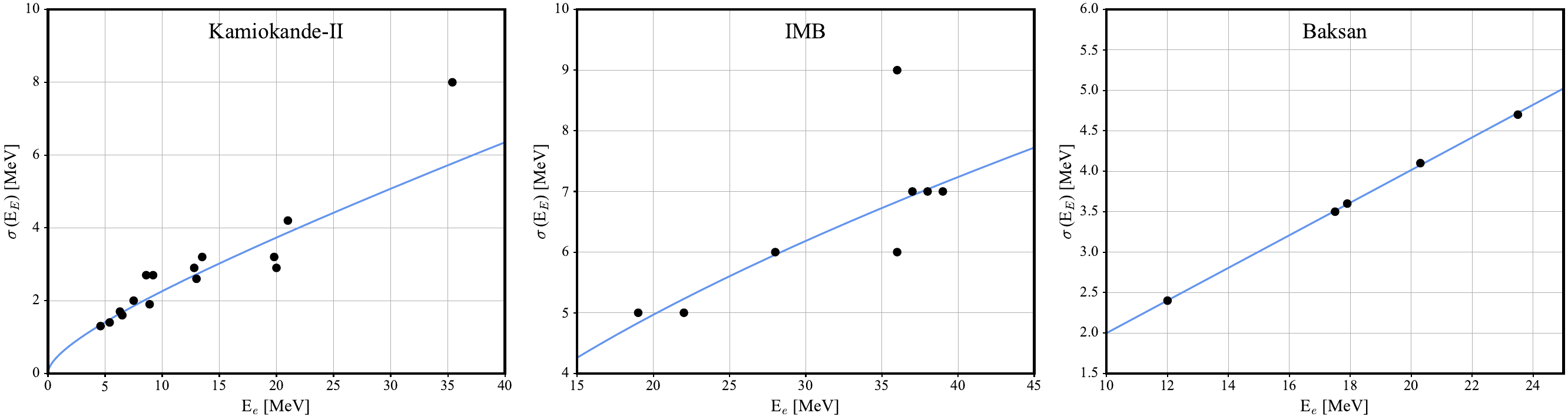}
    \caption{\emph{Parametrization of the uncertainty $\sigma(E_e)$ for each detector (Kamiokande-II, IMB, and Baksan) as a function of the positron energy $E_e$. %The uncertainty is modeled using the expression \(\sigma(E_e) = \sigma_{\text{stat}} \cdot \sqrt{E / (10 \, \text{MeV})} + \sigma_{\text{syst}} \cdot E / ( 10 \, \text{MeV})\), where $\sigma_{\text{stat}}$ and $\sigma_{\text{syst}}$, This parametrization is derived from the reported uncertainty values for the measured events and is be used in the kernel function which models the detector response.
    }}
    \label{fig:sigma_vs_energy}
\end{figure}

In figure~\ref{fig:sigma_vs_energy} we show the best fit \(\sigma(E_e)\) curves (solid lines) superimposed on the points provided by Kamiokande-II, IMB and Baksan. The corresponding resolution functions for each experiment are collected in table~\ref{tab:efficiency}. 

\begin{figure}[t!]
    \centering
    \includegraphics[width=1\linewidth]{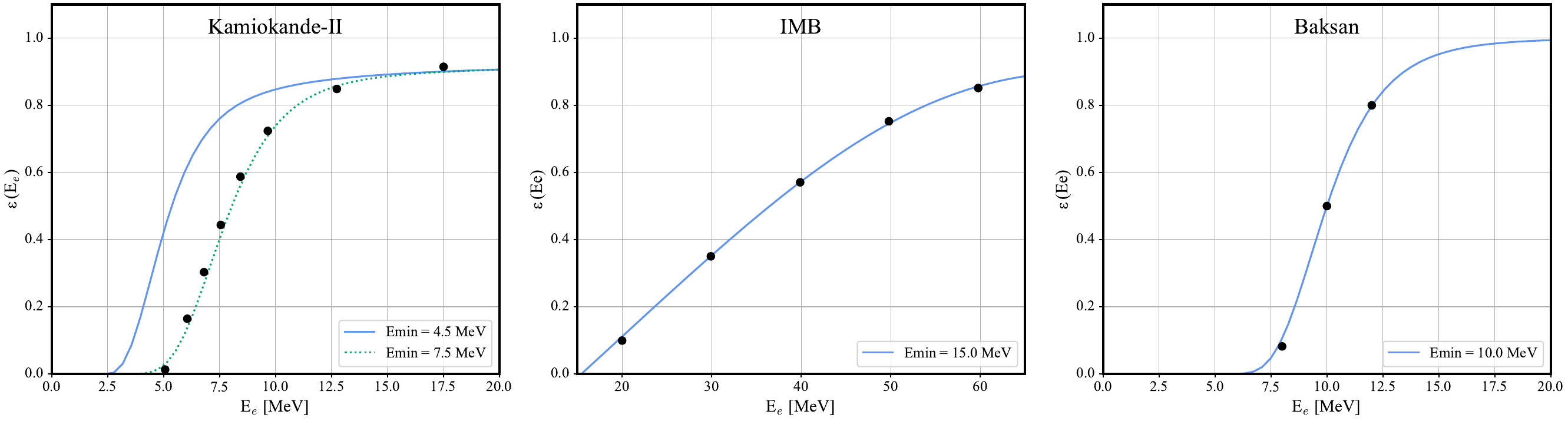}
    \caption{\emph{Detection efficiency $\varepsilon(E_e)$ as a function of electron energy $E_e$ for the three detectors: Kamiokande-II, IMB, and Baksan. The parametrization of the efficiency curves 
    %are parametrized
    is based on detector-specific acceptance thresholds, $E_{\min}$, as indicated in the legend. Black points represent the reported efficiency values from the detectors. 
    The continuous lines show the efficiency curves that we use for the analysis.}}
\label{fig:epsilon_vs_energy}
\end{figure}

\begin{comment} % Commento la tabella perché va bene quella con le parametrizzazioni intere a parer mio, questa è ridondante
\begin{table}[h!]
\centering
\begin{tabular}{@{}ccc@{}}
\toprule
\textbf{Experiment} & \(\sigma_{\mathrm{stat}}\) \,[MeV\(^{1/2}\)] & \(\sigma_{\mathrm{syst}}\) \\ \midrule
Kam-II & \(1.3 \,\pm\, 0.3\) & \(0.9 \,\pm\, 0.3\) \\
IMB    & \(3.3 \,\pm\, 1.2\) & \(0.2 \,\pm\, 0.7\) \\
Baksan & \(-0.03 \,\pm\, 0.05\) & \(2.0 \,\pm\, 0.1\) \\
\bottomrule
\end{tabular}
\caption{\emph{Fit values of \(\sigma_{\mathrm{stat}}\) and \(\sigma_{\mathrm{syst}}\) for the energy-resolution parameterization in eq.~\eqref{eq:sigma_parameterization}.}}
\label{tab:param_values}
\end{table}
\end{comment}

\paragraph*{Intrinsic and total efficiency:}
The total probability that a positron of true energy \(E_e\) is recorded as an event includes two key factors. The first is the threshold cut \(E_{\min}\), which is the minimum energy to admit an event in the analysis.  We adopt \(E_\min = 4.5\) MeV for Kamiokande-II in agreement with \cite{Vissani:2014doa}. We use all the events observed from SN1987A in the analysis, rather than excluding some by attributing them \emph{a priori} to background processes; in particular, we also consider K13, K14, K15, and K16 and the events recorded by Baksan.
 The second factor is the intrinsic efficiency \(\eta(E_e)\), which represents the fraction of events passing trigger, reconstruction, and identification, excluding resolution effects. While \(\eta(E_e)\) can be set to 1 in scintillators such as Baksan due to their high light yield, this is not the case for Cherenkov detectors.

\noindent To capture these effects together, we write the detector efficiency as in 
eq.~\eqref{eq:epsilonexpression}.
%\(\varepsilon(E_e)\) as
%\begin{equation}
%\varepsilon(E_e) \;=\; %\eta(E_e)\;\times\;\frac{1+\mathrm{erf%}\!\Bigl(\tfrac{E_e - E_{\min}}{%\sqrt{2}\,\sigma(E_e)}\Bigr)}{2}\,,
%\label{eq:efficiency_total}
%\end{equation}
We have adopted the parametrization and values for \(\eta(E_e)\) provided by ref.~\cite{Vissani:2014doa}. \footnote{We corrected a typo in \cite{Vissani:2014doa} concerning the parameterisation of $\eta$ for the IMB 
(eq.~14 there) and obtained slightly different coefficients, which however agree within the uncertainties.} Figure~\ref{fig:epsilon_vs_energy} shows the resulting \(\varepsilon(E_e)\) curves for Kamiokande-II, IMB, and Baksan, with black markers indicating the efficiencies reported by the respective collaborations at selected energies. For the reader's convenience, the full functions are provided in table~\ref{tab:efficiency}. %{\color{magenta} Our description of the efficiency  reproduces the value for $E_{\min}=5.6$ MeV reported in \cite{hiratasecondPhysRevLett.63.16}.}

\begin{table}
\renewcommand{\arraystretch}{1.9}
\begin{tabular}{|c|c|c|c|c|}
\hline  
 & \makecell{$E_{\min}$ \\ $[\unit{\MeV}]$} & \makecell{$\eta(E_e)$} & \makecell{$\sigma(E_e)$ \\ $[\unit{\MeV}]$} \\[5pt] 
\hline 
\text{Kamiokande-II} & 4.5 & $0.93\left[1-\left(\frac{0.2\, \unit{\MeV}}{E_e}\right)- \left(\frac{2.5\, \unit{\MeV}}{E_e}\right)^2 \right]$ & $1.27\left(\frac{E_e}{10\,\unit{MeV}}\right)^{1/2} + 1.0 \left(\frac{E_e}{10\, \unit{MeV}}\right)$ \\[5pt]
\hline 
\text{IMB} & 15 & \makecell{
    $0.379 \left(\frac{E_e}{15\,\unit{MeV}}-1\right)$ \\[3pt] 
    $-6\times 10^{-3} \left(\frac{E_e}{15\,\unit{MeV}}-1\right)^4$ \\[3pt] 
    $+ 10^{-3}\left(\frac{E_e}{15\,\unit{MeV}}-1\right)^5$ \\[5pt]}
& $3.3\left(\frac{E_e}{10\, \unit{\MeV}}\right)^{1/2} + 0.2\left(\frac{E_e}{10\,\unit{\MeV}}\right)$ \\[5pt]
\hline
\text{Baksan} & 10 & 1.0 & $2.0 \left(\frac{E_e}{10\,\unit{MeV}}\right)$ \\[3pt]
\hline
\end{tabular}
\caption{{\emph{Energy thresholds $E_{\min}$, intrinsic efficiencies $\eta(E_e)$ and energy standard deviation functions $\sigma(E_e)$ for Kamiokande-II, IMB and Baksan.} }}
\label{tab:efficiency}
\end{table}

\paragraph*{Angular Bias:}
Only for the case of IMB, the angular bias function is not trivial. At the time of SN1987A, IMB experienced a malfunctioning which affected 25\% of its readout channels, biasing the directional reconstruction. In our analysis, this is taken into account with a linear correction in efficiency $\zeta(\cos\theta)=1+0.1\cos \theta$, following the prescription of the IMB collaboration \cite{IMB:1988suc}.

%\subsubsection*{Further notes}
%We have tested minor variations in all the parametrizations and found that they do not critically affect the global fits or alter the key conclusions regarding the SN1987A neutrino signal. 

\section{Data analysis} 

\subsection{Methodology} 

There are two primary and distinct types of statistical analysis which are relevant
for the neutrino emission of SN1987A. One aims to identify, within a given class of parameterized models, which ones provide the best representation of the data (in particular, the best-fitting parameters with their confidence regions).  The other one aspire to assess how well  the observed data correspond to the fitted, assumed model,
%on assessing the fit of a specific model 
by means of goodness of fit (GOF) tests.
We have performed both analyses:
%; we report
%the results %obtained for the best fit and GOF tests will be 
%in section \ref{results}. 
%Regarding the 
%and underline here a few
%methodological aspects:
\begin{itemize}
 \item[1]    
the best-fit parameters of our model are obtained by minimizing  the $\chi^2$-function, as  detailed in section~\ref{s24}. 
In the statistical analysis, it is possible to apply a marginalization procedure by integrating over the parameters \cite{Loredo:2001rx} or, alternatively,   minimizing the  $\chi^2$. \footnote{This is equivalent to perform the  maximization of the likelihood function \cite{Pagliaroli:2008ur} (profile likelihood).} Although the former approach is preferable in principle, being theoretically more rigorous, it is significantly more demanding computationally.
Moreover,   no substantial discrepancy between the results given by the two methods is expected for the case of SN1987A \cite{Pagliaroli:2008ur}. We have therefore decided to adopt the latter procedure. 
\item[2]  The GOF tests are  quite standard and are detailed  in section~\ref{sec:gof}. 
\end{itemize}

%To achieve these objectives, 
%
We have chosen to perform all analyses in parallel using two different implementations and programming languages, Python and Mathematica. This decision enabled us to cross-check the results and optimize numerical computations by leveraging the combined strengths of both languages. Mathematica excels in symbolic computation and provides an extensive library of built-in functions that facilitate rapid prototyping and analytical derivations. Python, on the other hand, offers robust libraries for numerical analysis, data handling, and visualization, alongside excellent scalability and integration capabilities. By combining these strengths, we ensured that the codes developed in the two environments yield results that are highly consistent and comparable with remarkable precision.
%Qui va inserita una descrizione non solo statistica ma anche computazionale.
%Ho elencato alcuni punti
%Here we give the details of the sophisticated data analysis we have performed.
%Preparatory 

%Below we report the output of the statistical analysis, and in particular the best-fit parameters, together with their confidence regions and physical meaning.
%It is possible to apply marginalization by integrating over parameters \cite{Loredo:2001rx}, or use maximization of the likelihood function \cite{Pagliaroli:2008ur} (profile likelihood). While the former approach is preferable in principle, being theoretically more rigorous, it is computationally more demanding. In practice, however, no substantial discrepancies between the two methods has been claimed for the case of SN1987A. Our findings are consistent with those reported in the literature. 

%It is worth noting that, although marginalization is theoretically more rigorous, the comparison between the results of \cite{Loredo:2001rx} and \cite{Pagliaroli:2008ur} suggests that, in this specific case, it does not lead to significant differences. 
%
%The results of our analysis are discussed in section \,\ref{results}. 
Preliminary studies and results have already been presented \cite{di_risi_2024_13845875,oliviero2024}. %(talk Veronica e poster Giuseppe) doi

\subsection{Parameter determination
%Comparison with the observations
}\label{s24}
The parameters of the model are estimated by minimizing the $\chi^2$-function
\begin{equation}
    \chi^2=-2\sum_{d=k,i,b}\log{\mathcal{L}_d},
    \label{eq:chi_squared_function}
\end{equation}
where $\mathcal{L}_d$ is the unbinned poissonian likelihood of any detector ($k,i,b$ stand for Kamiokande-II, IMB and Baksan, respectively), defined as:
\begin{equation}
\label{eq:likelihood_function}
    \mathcal{L}_d=e^{-f_{\text{live}}\,N_{\text{tot}}}\times \prod_{j}^{N_d}e^{S(t_d+\delta t_j+\tau_d/2)\, \tau_d}\times\bigg[\frac{B_j}{2}+S(E_j,\cos{\theta_j,t_d+\delta t_j})\bigg],
\end{equation}
where $S$ is the signal rate, described above, and $B_j$ is the background given in table~\ref{tab:tab0}.
The index $j$ runs over the  events detected by each experiment. In our emission model  $t=0$ is the time of the arrival of the first {\em antineutrino} on Earth.
In the likelihood, we use the time $t_j$ of the detected event $j$,
%They  were not measured with sufficient precision, with the exception of IMB 
 %(respect to 
% the time of the arrival of the first antineutrino on Earth.)
 which can be written as 
\begin{equation}
t_j=t_d+\delta t_j\mbox{ for }d=k,i,b.
\end{equation}
Here $\delta t_j$ is the (measured) time difference between each detected event  and the first detected one. 
%In our emission model we set $t=0$ for the arrival of the first {\em antineutrino} on Earth. 
 The  parameters $t_{d}$  represent the  delay times between the {\em arrival}
 of the first $\bar \nu_e$ on Earth 
 ($t=0$)
 and the {\em detection} of the first event in each experiment. 
The delay times, first introduced in~\cite{Abbott:1987bm}, are 
estimated from the global fit. 
The energies $E_j$, the scattering angles $\theta_j$ and the $\delta t_j$ of each event are collected in table~\ref{tab:tab0}. 
For IMB, we set $\tau_d=0.035\,\unit{s}$ and $f_{\text{live}}=0.9055$  to take into account the dead-time and the muon contamination, while for both Kamiokande-II and 
Baksan $\tau_d=0 $ and $f_{\text{live}}=1$ ~\cite{Loredo:2001rx}.

In addition to the delay times, the model  depends on a set of 6 astrophysical parameters, identified and discussed in   section~\ref{s21}. These are  the time scales of the emission,
$t_{\text{max}}$, $\tau_a$ and  $\tau_c$, plus  $R_{ns}$, $\xi_{n0}$ and $T_0$.
%the radius of the neutron star and the initial 
%number of neutrons
%$R_{ns}$, $\xi_{n0}$, that parameterize the intensities of emission;
%$T_0$, the temperature that describes the initial average energy of antineutrinos, as detailed in Ref.~\cite{sym13101851}. 
We let them  vary in   wide ranges (our priors):
\begin{align}
     &\mbox{1~km}\leq R_{ns }\leq \mbox{100~km}&
    &0\leq \xi_{n0} \leq 40\% &
      &\mbox{2~MeV}\leq T_0 \leq \mbox{6~MeV}& \notag \\      &\mbox{1~s}\leq \tau_c \leq \mbox{10~s}&  
     &\mbox{100~ms}\leq \tau_a \leq \mbox{1~s}&  &\mbox{10~ms}\leq t_{\text{max}} \leq \mbox{200~ms}&    \\
        &\mbox{0}\leq  t_k\leq \mbox{0.5~s}&
        &\mbox{0}\leq  t_i\leq \mbox{0.5~s}& 
        &\mbox{0}\leq  t_b\leq \mbox{0.5~s}& \notag
\end{align}
In the following  these ranges will  be discussed  from a physical standpoint, compared  with  expected values in literature, and refined  in the course of the analysis.
Let us emphasize that 
$t_{\text{max}}$ is the time at which the antineutrino flux reaches its maximum value. Also,
the time-dependent functions  $\xi_n$ and $T_c$ 
%shown in subsection~\ref{s21}  
%defined in terms of the three time scales,and 
are respectively proportional to $\xi_{n0}$ and $T_0$. 

%, in Figure \ref{fig:luminosity}
%To illustrate clearly the continuity between the two phases, accretion and cooling, we plotted the anti-neutrinos luminosity using our model, see Figure \ref{fig:luminosity}. 
%obtained by continuously matching the average antineutrino energy in the two phases~\cite{Pagliaroli:2008ur}. 

We have estimated the confidence intervals by  scanning  the $\chi^2$-function around its minimum\footnote{This method allows to estimate the uncertainties when the parameters are (approximatively) Gaussian distributed around their best fit value. However, even when  the parameters are  distributed in a non-Gaussian way it provides often a reasonable approximation for the uncertainties.}, as described e.g.\ in~\cite{Lista:2016tva}. 
%   delay time %$t_d$ 
 Choosing a model parameter $p$ and letting the other model parameters vary in their prior interval, first we compute the profile function $\chi^2(p)$ from the  
eq.~\eqref{eq:chi_squared_function}, minimizing on the remaining variables; then we subtract the value in the minimum $p_{\mbox{\tiny best}}$, and in this way we obtain $\Delta \chi^2(p)=\chi^2(p) - \chi^2(p_{\mbox{\tiny best}})$.
 The best fit point of the parameters (corresponding to the maximum of the  likelihood function in eq.\,\eqref{eq:likelihood_function}), is now defined by the condition $\Delta \chi^2(p)=0$,  
while the  confidence level intervals are obtained by setting  $\Delta \chi^2=1, 6, 9$, corresponding to $1\sigma$, $2\sigma$ and $3\sigma$, respectively.

\begin{table}[t!]
    \centering
    \begin{tabular}{ccccccccccccc}
    \toprule
   &$R_{ns0}$&$\xi_{n0}$&$T_0$&$\tau_a$&$\tau_c$&$t_k$&$t_i$&$t_b$ \\
   $R_{ns0}$&1&0.60&-0.82&-0.059&-0.15&-0.0087&0.030&0.0054\\
   $\xi_{n0}$&0.60&1&-0.81&-0.26&0.30&-0.033&0.0097&0.0018\\
   $T_0$&-0.82&-0.81&1&-0.0094&-0.28&0.017&-0.013&-0.0026 \\
   $\tau_a$&-0.059&-0.26&-0.0094&1&0.071&0.015&-0.066&-0.022\\
   $\tau_c$&-0.15&0.30&-0.28&0.071&1&-0.0072&-0.013&0.00064\\
   $t_k$&-0.087&-0.033&0.017&0.015&-0.0072&1&-0.00013&-0.000016\\
   $t_i$&0.030&0.0097&-0.013&-0.066&-0.013&-0.00013&1&0.0019\\
   $t_b$&0.054&0.0018&-0.0026&-0.022&0.00064&-0.000016&0.0019&1\\
   \bottomrule
\end{tabular}
\caption{\emph{Correlation matrix of the parameters of the model when the rising time is fixed at $t_{\text{max}}=100\,\unit{ms}$. The correlation coefficients are evaluated under the hyphotesis of gaussian distributed parameters.}}
\label{tab:correlation_matrix}
\end{table}

\section{Results}
\label{results}

We start this section by summarizing  the results of our analysis (best fit parameters and their uncertainties), which we  discuss in detail later on. Let us preliminarily observe  that
we are not able to constrain simultaneously the rising time and the delay times. 
This can be attributed to the 
limited statistics from SN1987A, which does not allow to resolve appropriately the rising region. As a result  we fix the parameter  $t_{\text{max}}=100\,\unit{ms}$, as discussed in subsection~\ref{subsec:rising_time}. 

\subsection{Summary}
The minimum of the $\chi^2$-function in eq.\,\eqref{eq:chi_squared_function} is found at
\begin{align}
 & \notag R_{\text{ns}0} = (17.0)^{+0.7}_{-0.5}\,\unit{km}, \quad  \xi_{n0} = (0.018)^{+0.025}_{-0.011}, \qquad   T_0 = (4.6)^{+0.5}_{-0.4}\,\unit{MeV}\,, \\ 
 &t_{\text{max}} = 0.1 \,  \text{s}, \qquad \qquad \quad \tau_a = (0.52)^{+0.24}_{-0.15}\,  \text{s}, \qquad \quad \tau_c = (5.6)^{+1.8}_{-1.3} \,  \text{s}, \\ 
 &t_k = (0.035)^{+0.065}_{-0.024} \, \text{s}, \qquad t_i = (0.043)^{+0.102}_{-0.029} \, \text{s}, \qquad t_b = (0.054)^{+0.152}_{-0.041} \, \text{s}, \notag
\end{align}
  where $\chi^2=263.5$.
  Further information on the best-fit points and confidence intervals can be read from the  profile log-likelihood ($\Delta \chi^2$) functions.

   The correlation matrix computed as  the second derivatives of the $\chi^2$-function by means of the finite difference method is displayed in table~\ref{tab:correlation_matrix}.
  Let us observe that the astrophysical parameters and the delay times are not correlated; the largest correlations concern the first three parameters. One can therefore expect the timing determinations of the two emission phases to be little affected by the rest and therefore quite reliable.
  
   Figure~\ref{fig:luminosity} 
%It is possible to  
%clearly see 
clearly shows the smooth connection between the two phases of accretion and cooling.
\begin{figure}
    \centering
    \includegraphics[width=0.8\linewidth]{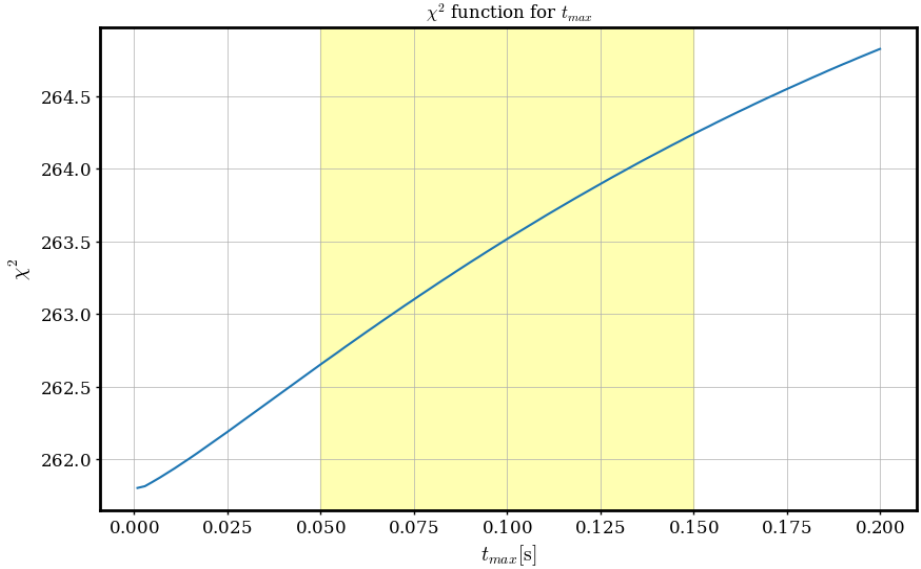}
    \caption{\emph{One-dimensional $\chi^2$-function of $t_{\text{max}}$. The yellow band indicates the region preferred by the simulations and adopted in the analysis.}
    %The prior on $t_{\text{max}}$ is $\left[0.01 -0.2\right]\,s$, as suggested in the literature. The $\Delta\chi^2$ variation is of few units over the prior and does not allow to discriminate a preferred value for $t_{\text{max}}$.
    }
    \label{fig:profile_t0}
\end{figure}

\subsection{Rising time}

%The matching function introduced in Ref.\,\cite{sym13101851} 

\label{subsec:rising_time}
Our model is characterized by an initial phase of antineutrino emission in which the luminosity increases reaching its peak at $t=t_{\text{max}}$
\cite{sym13101851}.  Simulations show that the expected    
$t_{\text{max}}$ is in the hundred milliseconds range \cite{Pagliaroli:2009qy,PhysRevD.80.087301,Kotake:2005zn, Marek:2007gr}. 
Let us investigate what insight the data from SN1987A provide. By letting $t_{\text{max}}$ vary in the interval $\left[0.01 - 0.2\right]\,\mbox{s}$, we minimize the function $\chi^2$. In figure~\ref{fig:profile_t0} it is shown that the resulting $\chi^2$ profile varies by only a few units over the interval--the data do not indicate a preferred value for $t_{\text{max}}$. Therefore, in the following analysis 
%we rely on a (tighter) theoretical prior, again 
 %shown in Fig.~\ref{fig:profile_t0}, and 
 we set $t_{\text{max}}$ at the central value $t_{\text{max}}=100\mbox{ms}$.
 %, and discuss the effect on inferences of varying this value in the interval $\left[50 - 150\right]\,\mbox{ms}$. 
 The alternative procedure of letting $t_{\text{max}}=0$, which would formally correspond to the assumptions of \cite{Loredo:2001rx, Pagliaroli:2008ur,DedinNeto:2023hhp}, is not supported by the simulations.

\begin{comment}
Our model features an initial phase of antineutrino emission where the luminosity increases reaching its peak at $t=t_{\text{max}}$
\cite{sym13101851}; consistently with the simulations we assume that  
$t_{\text{max}}$ is few hundreds of milliseconds \cite{Pagliaroli:2009qy,PhysRevD.80.087301,Kotake:2005zn, Marek:2007gr}. Let us discuss what we learn on $t_{\text{max}}$ directly from SN1987A data. Letting $t_{\text{max}}$ varies over the range $\left[0.01 - 0.2\right]\,\mbox{s}$, we minimize the $\chi^2$-function and normalize it by subtracting its global minimum. The resulting profile likelihood $\Delta\chi^2$ is displayed in Fig.\ref{fig:profile_t0}. It varies only few units over the prior interval, thus showing that the data do not indicate a preferred value for $t_{\text{max}}$. In the subsequent analysis we will fix $t_{\text{max}}=100\\mbox{ms}$, which is closer to theoretical expectations, and test the effect of varying this value in
the range  $\left[50 - 150\right]\,\mbox{ms}$.
%roughly corresponding to half of the variation of the one-dimensional $\chi^2$-function over the prior interval. 
\end{comment}

%Although the dynamics the dynamics of the initial increasing phase of the luminosity still presents a lot of open questions, it is expected that 

\subsection{Delay times}
\label{sec:delay_times}

Our analysis gives the  elapsed  time between the arrival of the first electron antineutrino and the detection of the first signal (delay time) for each experiment, as defined in section~\ref{s24}. They are different for the three experiments, not only because of the intrinsic detection properties but also because of the different spatial locations of the experiments. The latter contribution is {\em a priori} shorter than the time it takes for light to cross the Earth's diameter ($\lesssim 2 R_E/c=43$ms) and in the present case, it was several times smaller and therefore practically negligible. \footnote{SN1987A was in the deep southern sky and the detectors were all in the northern hemisphere instead.  The neutrinos reached first Kamiokande-II, then IMB and finally Baksan, after 7 and 10 ms respectively.}
Looking at the temporal distribution of the very first events, we established a reasonable prior $\left[0-0.5\right]\,\unit{s}$.

\begin{figure}[t!]
    \begin{center}
    % Prima immagine (a)
    \begin{subfigure}[b]{0.31\textwidth}
%$        \centering
        \includegraphics[width=\textwidth]{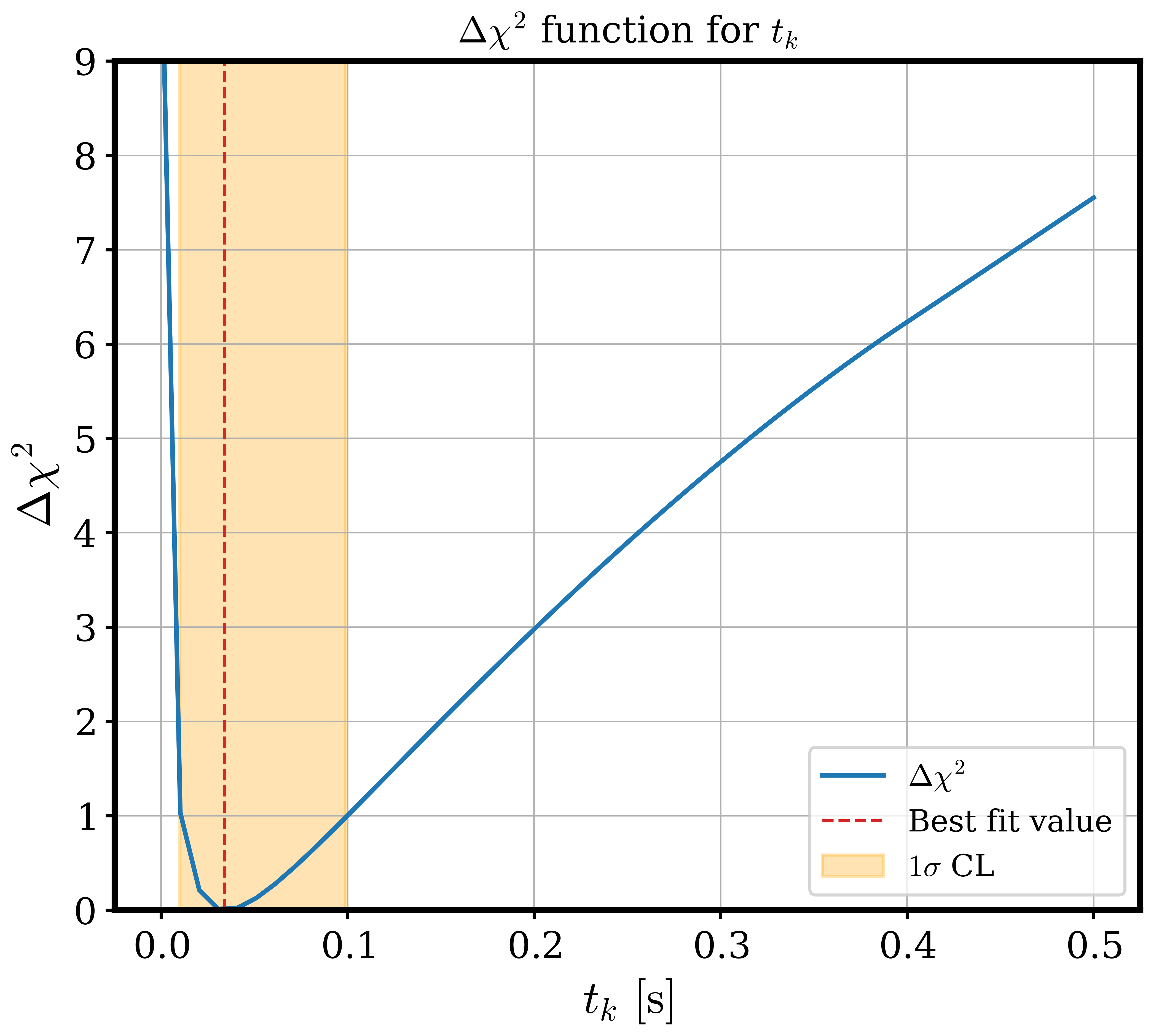}
        \caption{\emph{$\Delta \chi^2$-function of $t_{k}$}}
        \label{fig:profile_tk}
    \end{subfigure}
    %\quad
    % Seconda immagine (b)
    \begin{subfigure}[b]{0.31\textwidth}
      %  \centering
        \includegraphics[width=\textwidth]{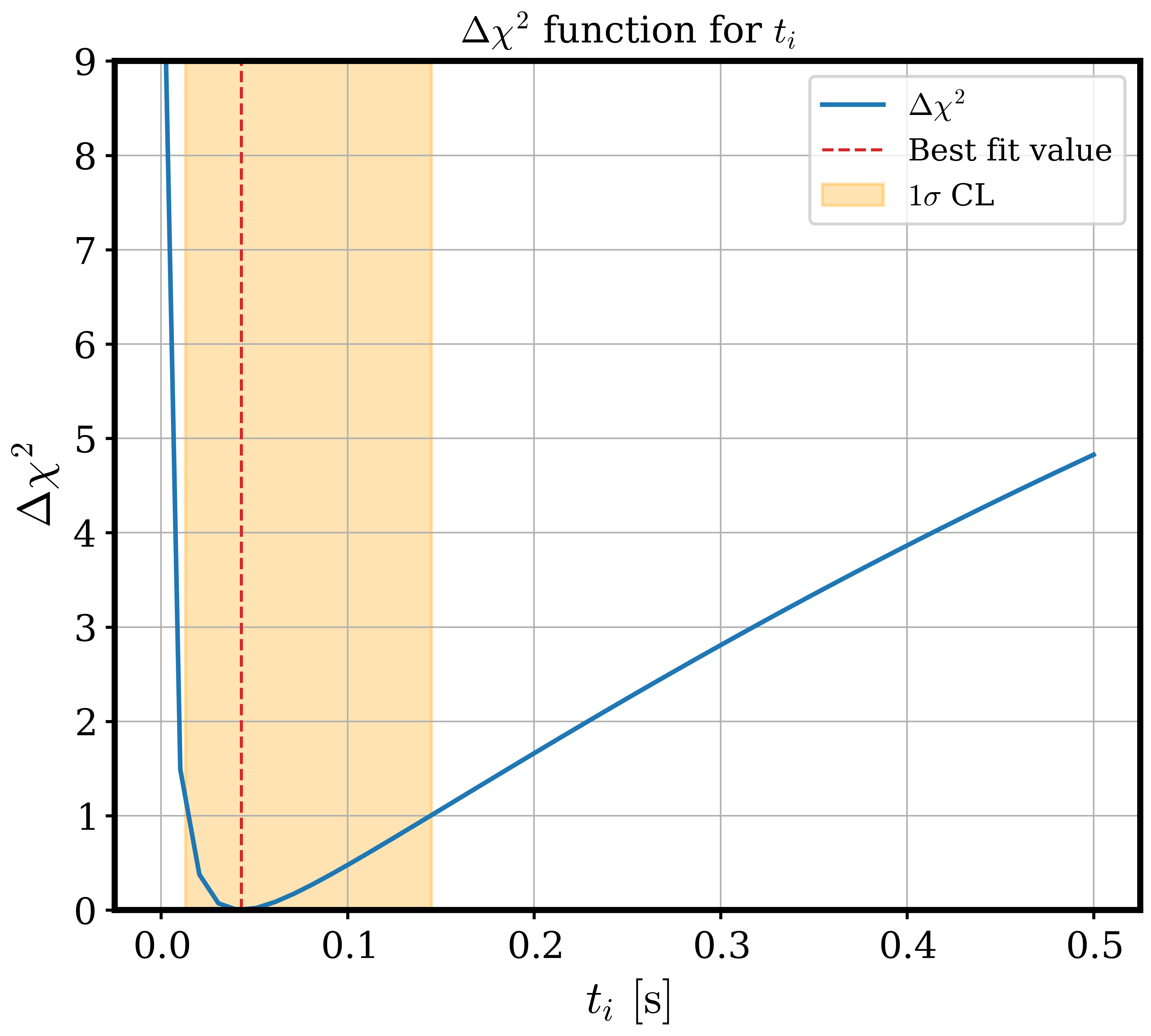}
        \caption{\emph{$\Delta \chi^2$-function of $ t_{i}$}}
        \label{fig:profile_tim}
    \end{subfigure}
   % \quad
    % Terza immagine (c)
    \begin{subfigure}[b]{0.31\textwidth}
      %  \centering
        \includegraphics[width=\textwidth]{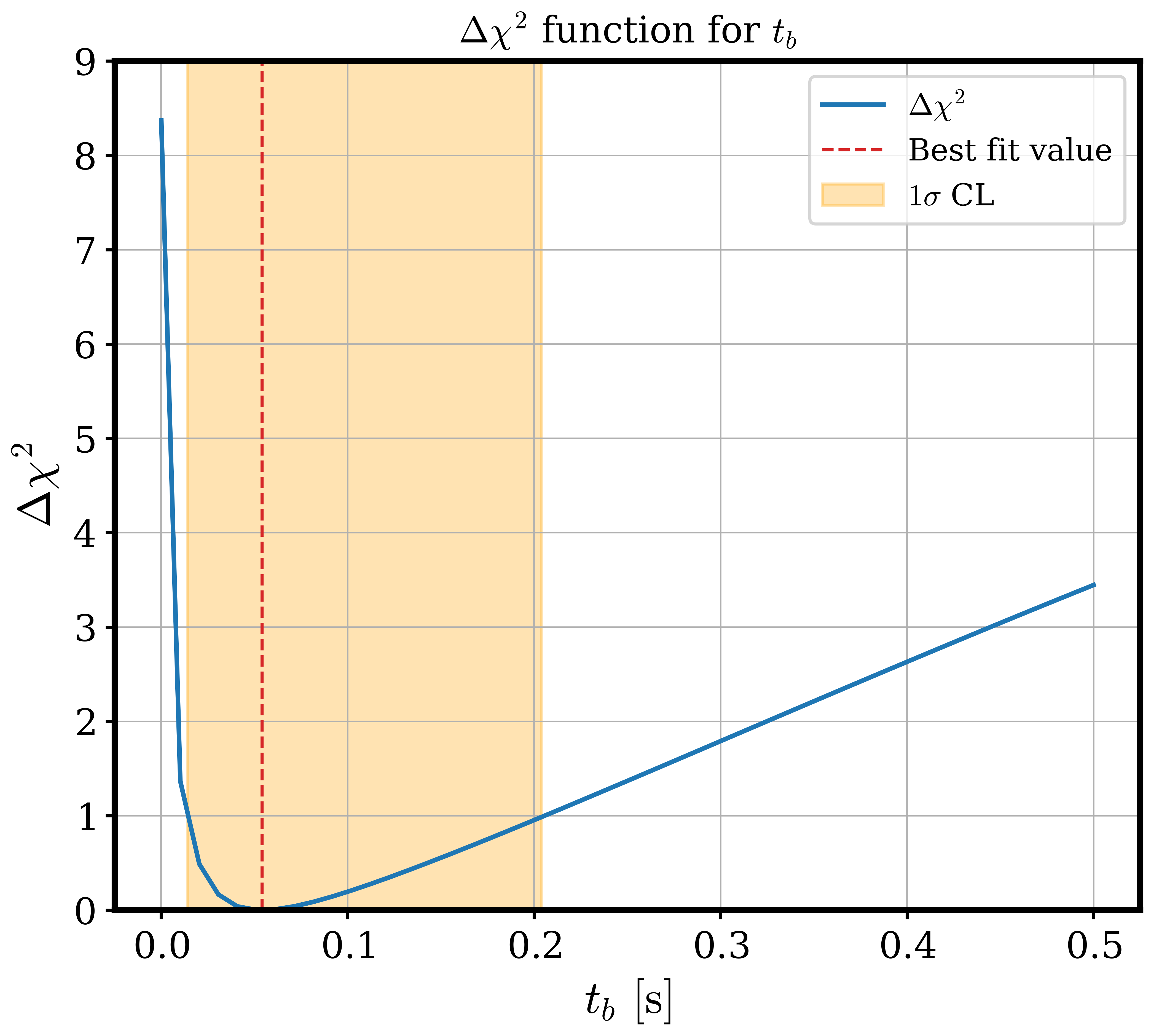}
        \caption{\emph{$\Delta \chi^2$-function of $ t_{b}$}}
        \label{fig:profile_tbk}
    \end{subfigure}
    \end{center}
    \caption{\em $\Delta \chi^2$-functions for the parameters $t_{k}$, $t_{i}$ and $t_{b}$ ($k,i,b$ stand for Kamiokande-II, IMB and Baksan, respectively). The red dashed line indicates the best fit values. The orange bands show the confidence interval at 1$\sigma$.}
    \label{fig:delays}
\end{figure}

 The best fit points with the 
%{\color{red} discussione piu' corposa}
uncertainties corresponding to the $1\sigma$-confidence 
intervals
are:
\begin{equation}\label{eq:delaytimefit}
      t_k=(0.035)^{+0.065}_{-0.024}\,\unit{\s},  \qquad  t_i=(0.043)^{+0.102}_{-0.029}\,\unit{\s}, \qquad   t_b=(0.054)^{+0.152}_{-0.041}\,\unit{\s} .
\end{equation}
The orange bands in figure~\ref{fig:delays}  represent the $1\sigma$-confidence intervals.
The adoption of a model of emission with a finite, initial, rising time for the neutrino luminosity represents
an improvement for the  estimates of the delay times in  previous analyses  where the luminosity features a nonphysical discontinuity at the onset time $t=0\,\unit{s}$ \cite{Loredo:2001rx,Pagliaroli:2008ur,DedinNeto:2023hhp}. 
Such safe description of luminosity together with an estimation of the delay times was also performed in \cite{Abbott:1987bm}, and, as far as we can compare, the results do not disagree. However, they consider the astrophysical model fixed,  do not take background into account and some events are excluded \textit{a priori} from the analysis. Differently, we prefer to include all the events registered in the $30\,\mbox{s}$ after the burst, thus using
all the %broad 
information provided by Kamiokande-II, IMB and Baksan.

\subsection{Accretion emission}
\label{subsec:accretion_emission}
\begin{figure}[t!]
    \begin{center}
    % Prima immagine (a)
    \begin{subfigure}[b]{0.33\textwidth}
        \includegraphics[width=\textwidth]{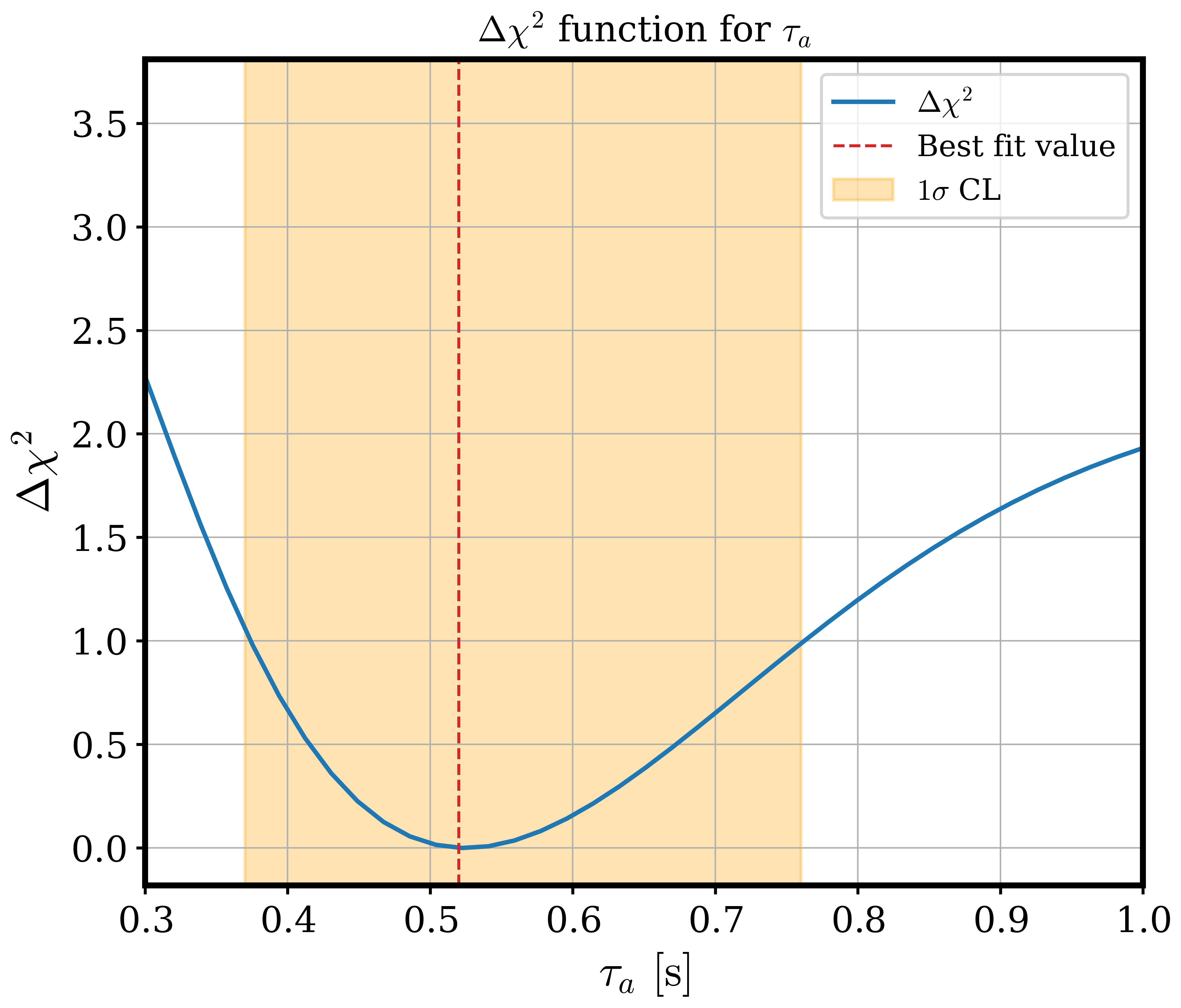}
        \caption{\emph{$\Delta \chi^2$-function of $\tau_a$}}
        \label{fig:profile_taua}
    \end{subfigure}
    %\quad
    % Seconda immagine (b)
    \begin{subfigure}[b]{0.31\textwidth}
        \includegraphics[width=\textwidth]{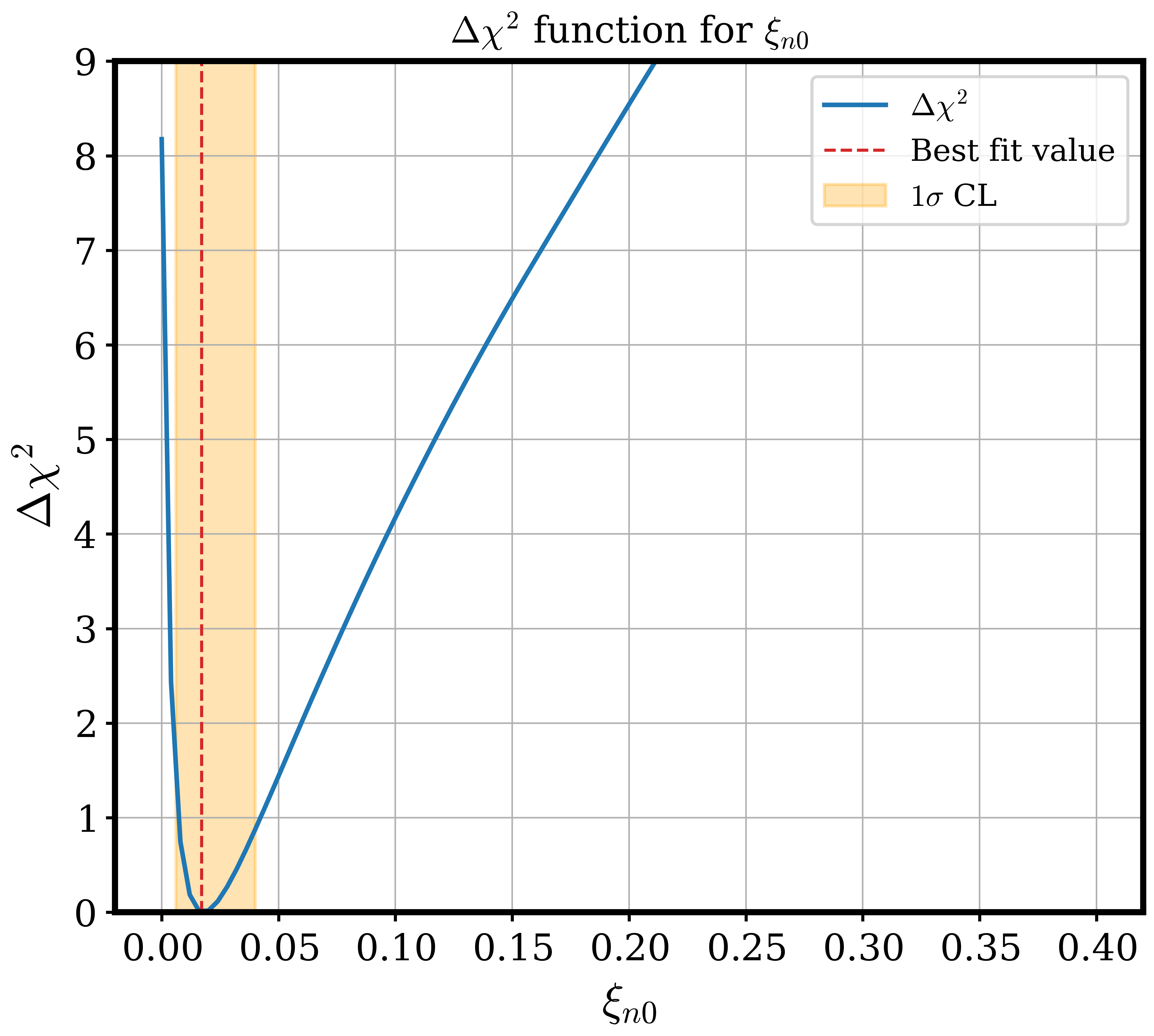}     \caption{\emph{$\Delta \chi^2$-function of $\xi_{n0}$}}
        \label{fig:profile_csi0}
    \end{subfigure}
    %\quad
    \begin{subfigure}[b]{0.31\textwidth}
        %\centering
        \includegraphics[width=\textwidth]{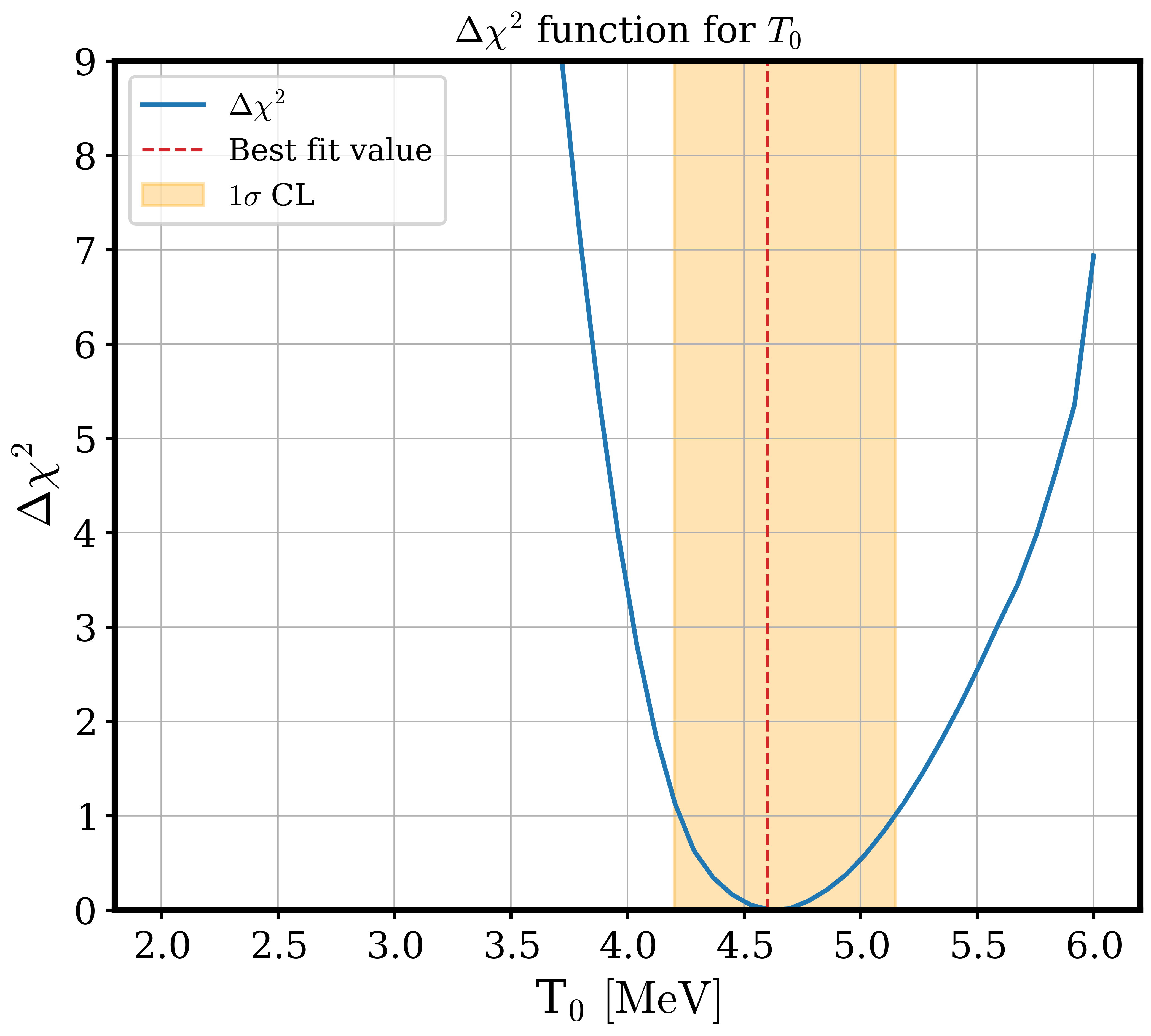}
        \caption{\emph{$\Delta \chi^2$-function of $T_0$}}
        \label{fig:profile_T0}
    \end{subfigure}
    \end{center}
 \caption{\em $\Delta \chi^2$-functions for $\tau_a$, $\xi_{n0}$ and $T_0$. The red dashed lines indicate the best fit value of the parameters. The orange bands show the confidence interval at 1$\sigma$.}
    \label{fig:taua_and_csi0_and_T0}
\end{figure}

The antineutrino emission in the  accretion phase depends on  $\xi_{n0}$, $T_0$ and $\tau_a$. From the procedure of minimization of the $\chi^2$-function we the obtain best-fit values
\begin{equation}
        \xi_{n0}=(0.018)^{+0.025}_{-0.011}  \qquad T_0=(4.6)^{+0.5}_{-0.4}\;\unit{\MeV }\qquad \tau_a=(0.52)^{+0.24}_{-0.15}\;\unit{\s }.
        \label{eq:accretion_best_fit}
\end{equation}
The $\Delta \chi^2$-functions for $\xi_{n0}$, $\tau_a$ and $T_0$ with $1\sigma$-confidence intervals (orange bands) are displayed in figure~\ref{fig:taua_and_csi0_and_T0}.
%, following the procedure described  in section\,\ref{sec:delay_times}.
The accretion time $\tau_a=0.52\,\unit{\s }$ is perfectly compatible with an initial accretion phase of a fraction of a second. We confirm also the un-physical feature for $\tau_a<0.3$~s, pointed out and discussed in section~3.3 of \cite{Pagliaroli:2008ur}. 
%Two quantitative outcomes deserve to be emphasized: 

%\noindent 1)
An interesting point to consider is %what is 
the statistical significance of this model compared to an emission model with only the cooling phase, % The model with only the cooling emission 
which can be formally obtained 
%from the one we use 
by setting $\xi_{n0}=0$. 
The variation of the minimum $\chi^2$ between the two models is $\Delta\chi^2\simeq 8.2$, a result which is somewhat weaker, but in substantial agreement with \cite{Pagliaroli:2008ur}, where $\Delta \chi^2=9.8-14.7$ for their models of emission. 

Our result corresponds to a significance of $99.8\%$ of the accretion hypothesis, evaluated by means of the likelihood ratio test \cite{Lista:2016tva,algeri2019searching} and considering that the model with no accretion (null hypothesis)  has one parameter fixed ($\xi_{n0}=0$). 
Alternatively, it could be assumed that the model without accretion actually has two fewer parameters
 ($\xi_{n0}$ and $\tau_a$), eventually obtaining a confidence level for the accretion of $99.2\,\%$.\footnote{The model with no accretion is obtained by putting $\xi_{n0}=0$, which is a borderline value for the model with accretion.  Consequently, the likelihood ratio $\Delta \chi^2$ is expected to be distributed as $f(x)=\frac{1}{2}\chi^2_m(x)+\frac{1}{2}\delta(0)$, using the notation adopted in \cite{algeri2019searching} where $m$ represents the difference between the number of degrees of freedom within the two models. The p-value  of the hypothesis with no accretion (null hypothesis) is $\int f(x)\dd x$ for $x > \Delta \chi^2$.}

\begin{comment}
However, since the existence of an accretion phase has been widely discussed in the last decades, we compute the statistical significance of our model with respect to a model of emission with only a cooling component by means of the likelihood-ratio test statistics. The model with only the cooling emission can be formally obtained by the one we use imposing $\xi_{n0}=0$. The variation of the minimum $\chi^2$ between the two is $\Delta\chi^2\simeq 8.2$, corresponding to a significance of $98.3-99.8\%$ of the accretion hypothesis, in accordance with Ref.\,\cite{Pagliaroli:2008ur}.
\end{comment}

 The fraction of neutrons involved in the emission,  
$\xi_{n0}=1.8\%$, despite its large uncertainty, suggests a scenario which is quite intriguing on physical grounds: during accretion,  $\bar\nu_e$ are produced mainly by a thin neutron atmosphere around the nascent proto-neutron star. This result represents a significant improvement with respect to previous analyses, where a large portion of the total mass of the outer core ($0.2-0.8 M_{\odot}$) was instead found to be 
needed at best fit point 
to take into account the observed spectrum of SN1987A \cite{Loredo:2001rx, Pagliaroli:2008ur}.

\subsection{Cooling emission}
\label{subsec:cooling_emission}
During the cooling phase, following the explosion, there is no longer significant accretion emission and the hot neutron star settles down by emitting neutrinos. Let us point-out that the  neutron star has long been sought without success, and this has caused a tension with theoretical expectations \cite{Shternin}. However, 
in 2015, very accurate simulations have showed that the hot neutron star is enveloped in a dense region of expanding gaseous remnants \cite{Orlando}.
 In addition to the indirect observational evidence  collected in the last years, direct evidence has recently been obtained \cite{Fransson:2024csf}.

In our model, the cooling temperature $T_c$ decreases over time according to eq. \eqref{cool:dep}, while the radius $R_{ns}$ of the neutron star remains constant. 
The spectrum of the emitted electronic antineutrinos is given in eq.\,\eqref{eq:spectrum_cooling}. The  parameters involved are the same temperature parameter $T_0$ that describes the accretion emission,  $R_{ns}$,  and $\tau_c$. 

\begin{figure}[t!]
    \begin{center}
    % Prima immagine (a)
    \begin{subfigure}[b]{0.4\textwidth}
        %\centering
        \includegraphics[width=\textwidth]{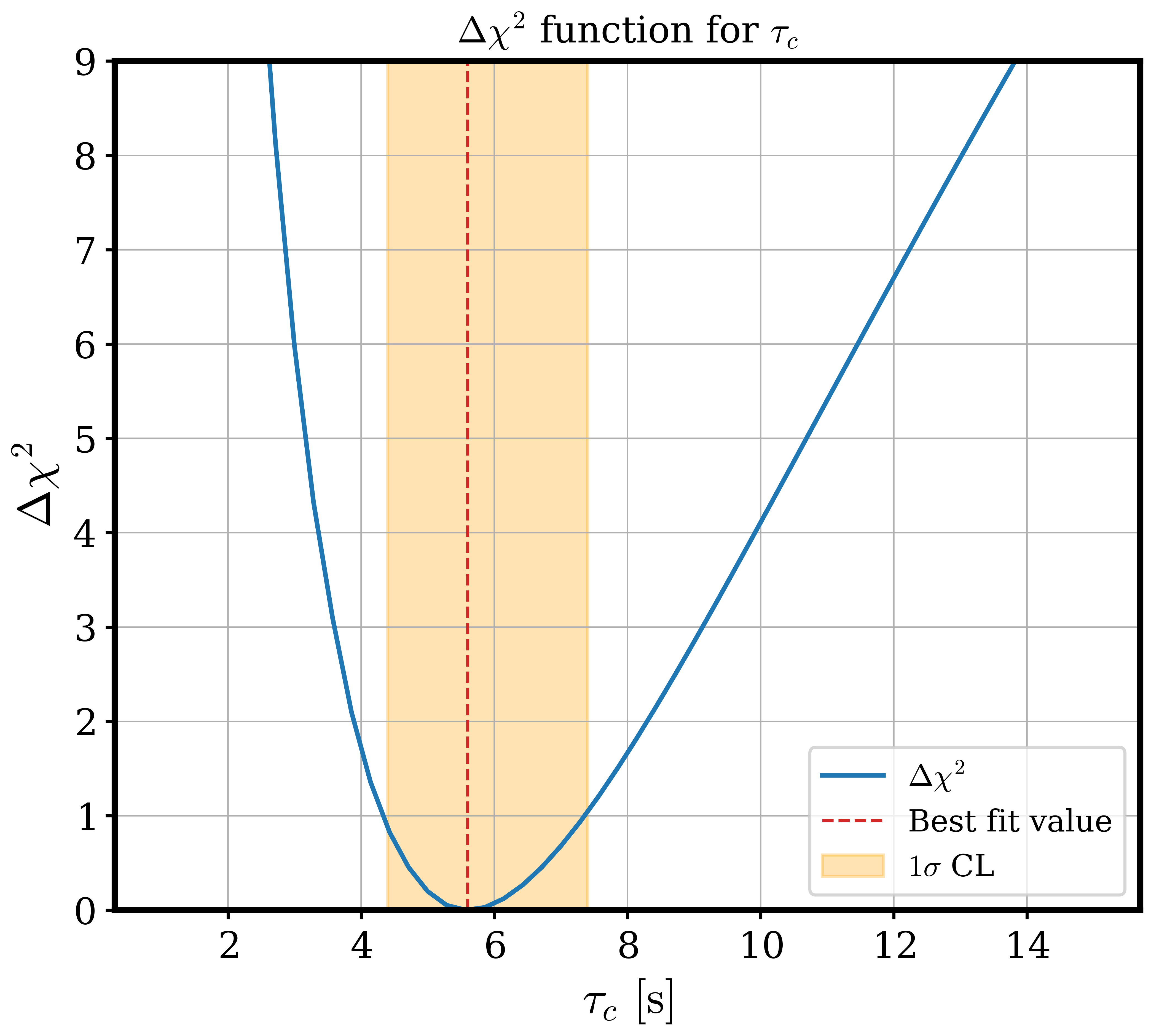}
        \caption{\emph{$\Delta \chi^2$-function of $\tau_c$}}
        \label{fig:profile_tauc}
    \end{subfigure}
    %\quad
    % Seconda immagine (b)
    \begin{subfigure}[b]{0.4\textwidth}
      %  \centering
        \includegraphics[width=\textwidth]{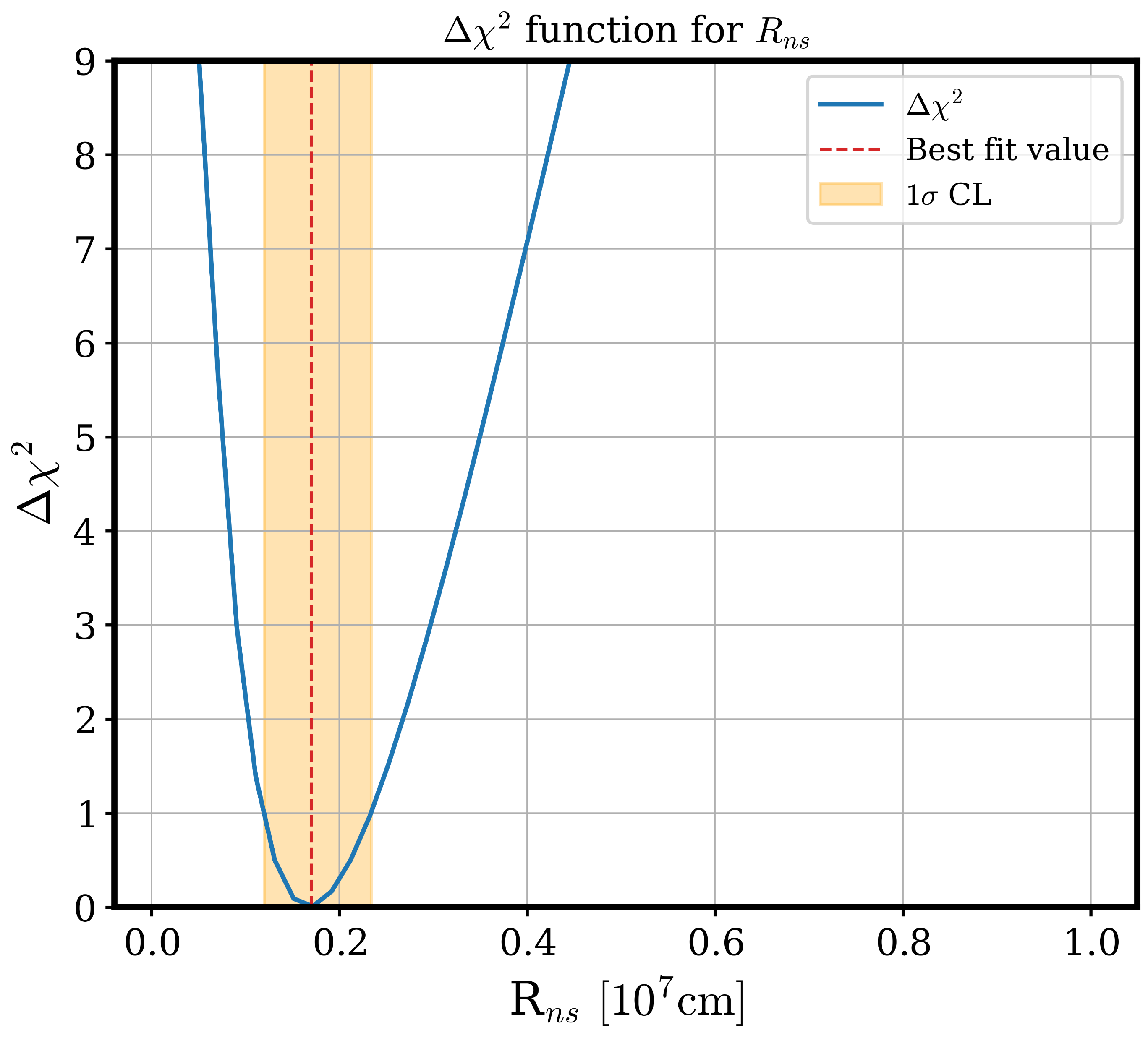}
        \caption{\emph{$\Delta \chi^2$-function of $ R_{ns}$}}
        \label{fig:profile_Rns}
    \end{subfigure}
    \end{center}
    \caption{\emph{$\Delta \chi^2$-functions for $\tau_c$ (left) and $R_{ns}$ (right). The red dashed lines indicate the best fit values. The orange bands show the confidence interval at 1$\sigma$.}}
    \label{fig:profile_tauc_Rns}
\end{figure}

%The temperature parameter $T_0$ has been presented in the previous section.  
The $\Delta \chi^2$-functions for $R_{ns}$ and $\tau_c$, together with the $1\sigma$-confidence intervals (orange bands), are shown in figure~\ref{fig:profile_tauc_Rns}. 
%The $1\sigma$-confidence intervals (orange bands in Fig.\,\ref{fig:profile_tauc_Rns}) are computed imposing the profile likelihood functions to be equal to $1$,  in accordance with the procedure outlined for the delay times in Sec.\,\ref{sec:delay_times}.
The numerical values for the radius and the cooling time at the best fit points are:
\begin{equation}
    R_{ns}=(17.0)^{+0.7}_{-0.5}\;\unit{\km},  \qquad \tau_c=(5.6)^{+1.8}_{-1.3}\;\unit{\s}.
    \label{eq:cooling_best_fit}
\end{equation}
%and reported in Eq.\eqref{eq:cooling_best_fit}.
The best fit value $R_{ns}=17\,\mbox{km}$ agrees with the expectation on the neutron star radius, improving the results of  previous analyses \cite{Loredo:2001rx,Pagliaroli:2008ur}--further %observational 
progress will come from its direct measurement. 
 The value for the cooling time $\tau_c=5.6\,\mbox{s}$ also fits well with the expected scenario of a long-standing thermal emission after the accretion \cite{Loredo:2001rx, Pagliaroli:2008ur}.
 
% \red{The cooling time $\tau_c$  controls the exponential decreasing on the neutrino luminosity as  $\sim 4\,\tau_c$: see Eqs.~\eqref{eq:Fcalligrafica} and \eqref{cool:dep} and discussion therein. Evidently, the same parameter rules the time of event accumulation.} 

Recently, improved simulations of the duration of the emission have been obtained, suggesting that $95\%$ of Kamiokande-II and IMB signal events are collected in $5-7\,\mbox{s}$, with some models accounting for durations up to $10\,\mbox{s}$ \cite{Fiorillo:2023frv}. Similarly, these simulations show that Baksan collects  $95\%$ of data in $4-10\,\mbox{s}$, but, differently from the analyses for Kamiokande-II and IMB, the background is included. 

%This agrees well with the results of our data analysis. 
At the best fit point  the number of events expected as  signal in our model  (no background) are
\begin{equation}
    N_{\text{KII}}=16.2\pm3.7 \qquad N_{\text{IMB}}=6.2\pm1.9 \qquad N_{\text{BNO}}=2.1\pm0.4,
\end{equation}
%We have evaluated the uncertainties by means of the correlation matrix in Tab.\ref{tab:correlation_matrix}.
%
The corresponding time windows to cumulate the $95\%$ of the whole set of events are
% \begin{subequations}
 \begin{eqnarray}
     \text{Kamiokande-II}\qquad 10.8\,\unit{s}&
\\
    \text{IMB}\quad\qquad\qquad 7.3\,\unit{s}&
\\
     \text{BNO}\quad\qquad\qquad 9.8\,\unit{s},&
\end{eqnarray}
 in perfect agreement with some  models of emission considered in ref.~\cite{Fiorillo:2023frv}, in particular the ones with the neutron star mass of $1.62 M_{\odot}$ and equations of state of the stellar medium without convection. Furthermore,  when uncertainties in astrophysical parameters are taken into account, 
 our results do not contradict most of the models in  tables VII and VIII of ref.~\cite{Fiorillo:2023frv}.

\subsection{Goodness of Fit}
\label{sec:gof}

While obtaining the best fit parameters is an essential step of an  accurate statistical analysis,  it is equally important to confirm that the model reliably reproduces the observed data from SN1987A. Noteworthy in passing is that such scrutiny
%statement 
should apply  not only to data analyses as the present one, but also to theoretical simulations of the neutrino flux emitted by SN1987A.

To ensure that our model is consistent with the observed data, we have verified its goodness of fit (GOF). 
We have chosen the Cramér–von Mises and Kolmogorov-Smirnov tests, which compare the empirical cumulative distribution function (ECDF) of a sample with the cumulative distribution function (CDF) of a specified theoretical distribution. The goal of the tests is to evaluate at which confidence level the %observed
data are consistent with the theoretical model.

\begin{table}[t!]
\centering
\begin{tabular}{|>{\raggedright\arraybackslash}p{3cm}|c|c|>{\centering\arraybackslash}p{3cm}|>{\centering\arraybackslash}p{3cm}|}
\hline
\textbf{p-values Cramér–von Mises test}  &  \textbf{Baksan} & \textbf{IMB}& \textbf{Kamiokande-II} (including all events)  &\textbf{Kamiokande-II} (excluding K13-K14-K15-K16)\\ \hline
\textbf{Rate} &  77.8 \% &  89.4\% & 60.9 \% &-\\ \hline
\textbf{Energy} & 55.0 \% & 10.9 \% & 18.3 \% & -\\ \hline
\textbf{Angle} &   N/A &  5.9\% & 38.0 \% & 7.0\,\% \\ \hline
\end{tabular}

\vspace{1em}

\begin{tabular}{|>{\raggedright\arraybackslash}p{3cm}|c|c|>{\centering\arraybackslash}p{3cm}|>{\centering\arraybackslash}p{3cm}|}
\hline
\textbf{p-values Kolmogorov-Smirnov test } & \textbf{Baksan} & \textbf{IMB}& \textbf{Kamiokande-II} (including all events) &\textbf{Kamiokande-II} (excluding K13-K14-K15-K16) \\ \hline
\textbf{Rate}  & 77.9 \% & 91.1 \% &  45.6 \% & -\\ \hline
\textbf{Energy} &  57.3 \% & 4.4 \% &  15.9 \% & -\\ \hline
\textbf{Angle} & N/A &  11.9\% &  49.0 \% & 11.9\,\%  \\ \hline
\end{tabular}
\caption{\emph{p-values from the comparison of the theoretical time, energy and angle cumulative distributions (CDF) with the empirical cumulative distributions (ECDF) observed by Kamiokande-II, Baksan and IMB. 
The p-values are computed using the Cramér–von
Mises and the Kolmogorov-Smirnov tests. 
Baksan does not measure the direction of the events, so there is no information on the empirical angular distribution. 
\label{table:gof_cramer}}}
\end{table}

From the observed signal, differential in energy, time and $\cos \theta$ in eq.\,\eqref{eq:differential_observed_signal}, and the background (see table 2 in \cite{Vissani:2014doa}), we compute the theoretical energy, temporal and angular CDF for Kamiokande-II, Baksan, and IMB. The benchmark values for the parameters in the signal are the results of our best fit calculations.
%
%The Cramér–von Mises test was then performed using the SN1987A neutrino events observed by the Kamiokande-II, IMB, and Baksan experiments. 
The ECDF are built %on 
with the data 
reported %collected 
in table~\ref{tab:tab0}. For the comparison with the theoretical angular distribution of Kamiokande-II we perform two analyses: 
\begin{itemize}
\item[1]
we exclude the events K13, K14, K15, K16, which are usually assumed to be background. 
%We stress that the threshold analysis $E_{min}$ must be changed to $7.5\,\unit{MeV}$, Coherently;
Consistently with this assumption 
%We stress that
we change the threshold $E_{\min}$ to $7.5\,\unit{MeV}$; 
\item[2]
we include all events, hence also the events K13, K14, K15, K16, exploiting the information on the scattering angles in ref.~\cite{Krivoruchenko:1988zg}.
\end{itemize}
The confidence level resulting  from these tests, expressed by the p-values, are reported in table~\ref{table:gof_cramer}. 

The values indicated in table~\ref{table:gof_cramer} show that the flux obtained from our statistical analysis describes the data well, and this is particularly true for its temporal distribution. However, we note  some tension between: 1)~the energy spectra of Kamiokande-II (‘cooler’) and IMB (‘hotter’); 2)~the angular distribution expected from the inverse beta decay and the observed one. These are well-known features, much discussed in the literature \cite{Malgin:1998wm, Krivoruchenko:1988zg,Costantini:2004ry}, 
  but without a clear conclusion.
  \footnote{There have been attempts to explain the discrepancy of the angular distribution assuming that some events are due to elastic scattering (ES) rather than inverse beta decay. 
For a discussion of the best candidate ES, the first Kamiokande-II event, see \cite{Costantini:2004ry,Vissani:2008wbj}.
It is not very likely that ES events are present in the data of IMB, since we expect (due to the neutrino in the final state) the visible energy to be small, and especially since none of them is well aligned with the supernova.}
%
%For what concerns our analysis, in light of the quantitative indicators shown here, 
Our %tentative 
interpretation is that
 the observed deviation from the theoretical angular distribution can be reasonably attributed to the particular data set.

\begin{figure}
    \centering
\includegraphics[trim={80 0 0 0},
width=1.1\linewidth]{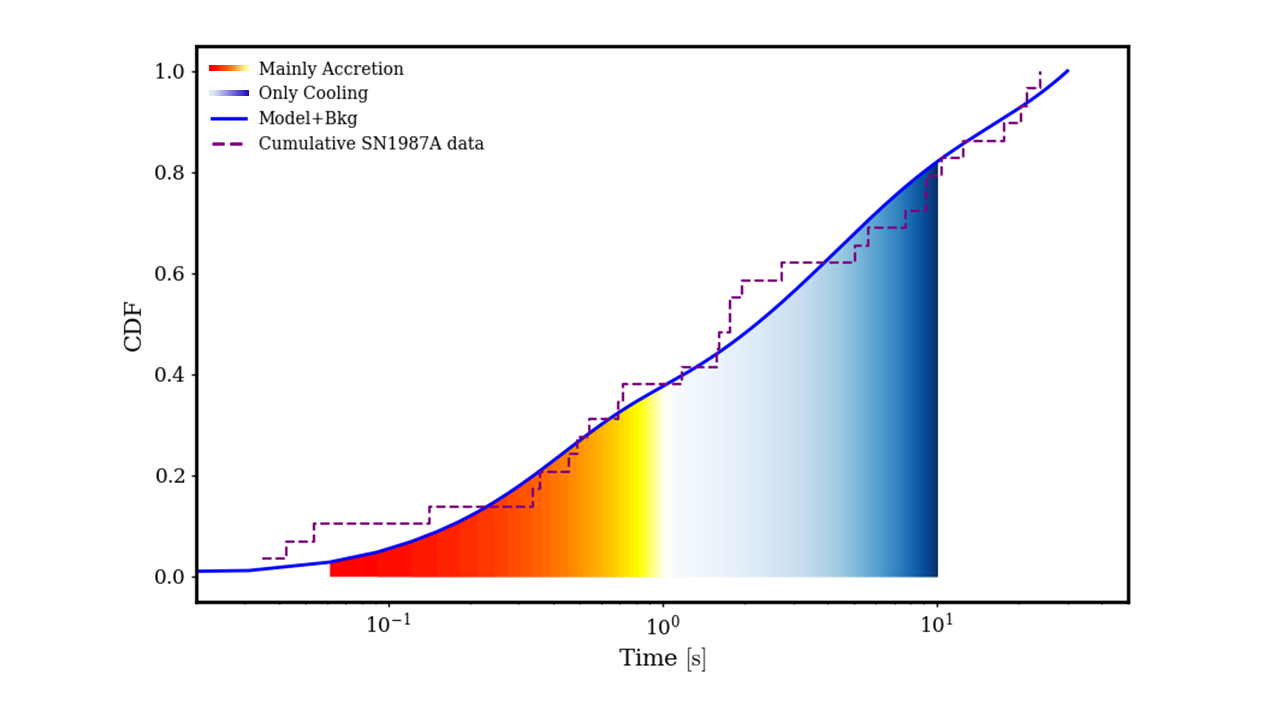}
    \vskip-3mm
    \caption{\emph{Comparison between the temporal CDF of our model and the ECDF from the data of all three experiments, including delays. The color scale  is the same of 
    %in agreement with 
    %that in
    figure~\ref{fig:luminosity}, and again it helps to discriminate the two emission phases. 
    Note the use of the
    %We use  a 
    logarithmic time scale. %to emphasize the temporal features. 
    }}
    \label{fig:combinedcdf}
\end{figure}

As a matter of fact, from the last two columns of table~\ref{table:gof_cramer} we see    that, when the events K13, K14, K15 and K16  are included (rather than excluded {\em a priori} from the analysis),  the GOF  of the angular distribution of Kamiokande-II improves. 

Finally, let us discuss the GOF of 
the entire data set: 
\begin{enumerate}
\item energy distribution: we find a significantly better agreement with the %theoretical 
energy CDF, as quantified by the  Kolmogorov-Smirnov p-value $50.7\,\%$ and Cramér–von
Mises p-value~$77.7\,\%$;
\item time distribution: 
 employing  the additional information on the delay times obtained from our analysis, we are able to compare the %theoretical 
 temporal CDF, which includes all three experiments, with the data. The result is displayed in figure~\ref{fig:combinedcdf}. The excellent agreement is confirmed by the goodness-of-fit tests we have already used for each experiment: the Kolmogorov-Smirnov p-value is $82.8\,\%$, while the Cramér–von
Mises p-value is $88.4\,\%$;
\item angle distribution: the %theoretical 
CDF 
 has a poorer agreement with the available empirical distributions. 
 %have less good agreement with the empirical distributions available.
 The Kolmogorov-Smirnov p-value is $5.5\,\%$, while the Cramér–von
Mises p-value is $2.0\,\%$, which is
%a bit 
less problematic than previous findings in  literature 
\cite{Malgin:1998wm, Krivoruchenko:1988zg,Costantini:2004ry} although not entirely satisfactory. 
\end{enumerate}
 %  {\color{magenta}
%Summarizing, our model describes reasonably well the data collected from SN1987A.}

\section{Conclusions}

We have  
described the 
 neutrino emission from SN1987, by considering the time, energy, and scattering angle of the events.  We have performed a refined analysis of the data adopting a parameterized model  \cite{sym13101851}.
Parametric models, characterized by physically motivated parameters, are straightforward to understand and flexible enough to accommodate future observations. 
Our model improves on the previous ones in literature by including  an  initial {\em finite} rise time, not yet observed because of the limited number of events but in line  with the expectations.
It accounts for a smooth  evolution between the accretion and cooling phases of luminosity, i.e., it provides a continuous and regular curve, similar to those indicated by the simulations.
We  have assumed that all detected neutrino events come from inverse beta decay, which is the most relevant channel for non-thermal processes.
For the IBD cross section we have adopted the most recent and accurate calculation to date \cite{Ricciardi:2022pru}.

In our analysis, we have exploited the energy, time and direction of arrival of the events as provided  by Kamiokande-II \cite{hirataPhysRevLett.58.1490,PhysRevD.38.448},  IMB \cite{PhysRevLett.58.1494,PhysRevD.37.3361} and  Baksan \cite{ALEXEYEV1988209}, Furthermore, we did not classify the individual events as signal or background {\em a priori} but analyzed them keeping into account all available information.\footnote{Indeed, in the absence of the identification of the  neutron, associated to the positron in the IBD reaction, no single event can be attributed with absolute certainty to the supernova, and useful statements refer to sufficiently large subsets of data, ideally the entire data set.} Throughout the analysis, we  have used the information on the  expected background in the experiments, considering their specific  features and biases.
We have recovered 
 the intrinsic (or hardware) efficiency of each experiment as a function of energy by the best available information and reasonable extrapolation. 
 %, see appendix~\ref{app:2}.
%This is necessary for  the  correct use of total and intrinsic efficiencies in the construction of the likelihood.
The total efficiencies (provided by the experiments) define the total expected number of events, while the intrinsic efficiencies $\eta$  and the resolution functions $\sigma$
are needed to model the triple differential emission spectra.

We have performed a  statistical analysis which is in several respects  more refined and precise than earlier efforts in literature.
The result  are the best fit values of 9 parameters, describing astrophysics and detector response, 
together with their confidence regions. Let us underline the main results:
\begin{itemize}[label=--]
 \item we have obtained the first accurate estimates  of the initial rising time $t_{\text{max}}$, and of the experimental delay times. We have showed that data do not indicate a preferred value for $t_{\text{max}}$, that we have set as a prior at 
$t_{\text{max}}=100\mbox{ms}$.
The delay times can be calculated within uncertainties of 100~ms.
\item We   give accurate values for the duration of the accretion and cooling phases, in agreement with current literature.  We find that the statistical significance of considering both the two primary phases, with respect to only considering the cooling phase, is 98.8-99.8$\%$.
\item The best fit values of the remaining parameters are also in accordance with current simulations of SN explosion. 
\item A particularly intriguing result regards the value of 
the fraction of neutrons involved in the emission, which for the first time align closely with expectations. Despite its large uncertainty, the value $\xi_{n0}=1.8\%$ suggests that during accretion,  $\bar\nu_e$ are produced by a  neutron atmosphere around the nascent proto-neutron star thinner than expected from previous analyses \cite{Loredo:2001rx, Pagliaroli:2008ur}.
\item Other interesting results pertain to 
the number of events expected as  signal,
 together with their uncertainties, and 
the  time windows to cumulate the $95\%$  of the events, that we have compared with the expectations. 
\end{itemize}

  We have cross-checked  all analyses by always using two different programming languages, Python and Mathematica. 
We have assessed the suitability 
%the adeguacy 
of our model at the best fit values  through goodness-of-fit tests,   performed on the temporal,
energy and angular distributions.
We have used the Cramér–von Mises and Kolmogorov-Smirnov tests, and compared 
 the empirical cumulative distribution function  of a sample with the cumulative distribution function  of a specified theoretical distribution (see figure~\ref{fig:combinedcdf}). We find the   description of the flux quite reliable,   particularly the temporal distribution.

\section{Acknowledgments}
  V.d.R. would like to thank the Quantum Theory Center (QTC) at the Danish
Institute for Advanced Study and IMADA of the University of Southern Denmark for the hospitality during the
completion of this work. G.R. acknowledges the support of the research projects  ENP (Exploring New Physics), funded by INFN (Istituto Nazionale
di Fisica Nucleare).
The work of F.V. is partially supported by the grant 2022E2J4RK {\em PANTHEON: Perspectives in Astroparticle and Neutrino THEory with Old and New messengers,}
PRIN 2022, funded by the Italian Ministero dell’Universit\`a e Ricerca (MUR) \& European Union – Next Generation EU.
\newpage

\flushbottom
\newpage
\bibliographystyle{JHEP}
\bibliography{biblio}
\end{document}